\documentclass[preprintnumbers,amsmath,amssymb,twocolumn,superscriptaddress]{revtex4}

\usepackage{CJK}
\usepackage{graphicx}
\usepackage{subfigure}
\usepackage{dcolumn}
\usepackage{bm}
\usepackage{booktabs}
\usepackage{threeparttable}
\usepackage{multirow}
\usepackage{slashed}
\usepackage{array}
\usepackage{makecell}
\usepackage{enumerate}
\usepackage{textcomp}
\usepackage[colorlinks,
            citecolor=blue,
            anchorcolor=red,
            menucolor=red,
            linkcolor=red,
            filecolor=red,
            runcolor=red,
            urlcolor=blue,
            frenchlinks=red]{hyperref}

\newcommand{\PreserveBackslash}[1]{\let\temp=\\#1\let\\=\temp}
\newcolumntype{C}[1]{>{\PreserveBackslash\centering}p{#1}}
\newcolumntype{R}[1]{>{\PreserveBackslash\raggedleft}p{#1}}
\newcolumntype{L}[1]{>{\PreserveBackslash\raggedright}p{#1}}
\newcommand\keyword[1]{\textbf{Keywords}: #1}

\begin{document}
\begin{CJK*}{GBK}{song}
\title{Inelastic heavy quarkonium photoproduction in $p$-$p$ and $Pb$-$Pb$ collisions at LHC energies}
%{\CJKfamily{gbsn}
\author{Zhi-Lei Ma}
%(马智磊)
\affiliation {Department of Physics, Yunnan University, Kunming 650091, China}
\affiliation {Department of Astronomy, Key Laboratory of Astroparticle Physics of Yunnan Province, Yunnan University, Kunming 650091, China}

\author{Zhun Lu}
%\email{zhunlu@seu.edu.cn}
\affiliation {School of Physics, Southeast University, Nanjing 211189, China}
%(吕准)

\author{Hao Liu}
%(刘浩)
\affiliation {Department of Physics, Yunnan University, Kunming 650091, China}

\author{Li Zhang}
\email{lizhang@ynu.edu.cn}
\affiliation {Department of Astronomy, Key Laboratory of Astroparticle Physics of Yunnan Province, Yunnan University, Kunming 650091, China}
%(张力)
\date{\today}

\begin{abstract}
\textbf{Abstract:} We study the inelastic charmonium ($J/\psi$, $\psi(2S)$) and bottomonium ($\Upsilon(nS)$) photoproduction and fragmentation processes in $p$-$p$ and $Pb$-$Pb$ collisions at LHC energies, where the ultra-incoherent photon emission is included.
In the framework of the NRQCD factorization approach, an exact treatment is developed which recovers Weizs\"{a}cker-Williams approximation (WWA) near the region $Q^{2}\sim0$, where the methods of Martin-Ryskin and BCCKL are used to avoid double counting.
We calculate the $Q^{2}$, $y$, $z$, $\sqrt{s}$, $p_{T}$ dependent and the total cross sections.
It turns out that the inelastic photoproduction and fragmentation processes provide valuable contributions to the heavy quarkonium production, especially in the large $p_{T}$ regions.
While the relative contribution of ultra-incoherent photon channel is very important,  which rapidly increases along with the growing quarkonium mass, and begins to dominate the photoproduction processes at large $p_{T}$ ranges.
Moreover, we obtain the complete validity scopes of WWA in inelastic heavy quarkonium photoproduction in heavy-ion collisions.
WWA has a much higher accuracy at high energies and in $Pb$-$Pb$ collisions.
The existing photon spectra are generally derived beyond the applicable scopes of WWA, and the double counting exists when the different channels are considered simultaneously.\\
\keyword{inelastic photoproduction, heavy quarkonium, exact treatment, WWA, NRQCD}
\end{abstract}

%\pacs{25.75.Cj, 25.20.Lj, 12.39.St, 12.38.Mh}
%\footnote{Corresponding author}
%\begin{keywords}
%inelastic photoproduction, heavy quarkonium, exact treatment, WWA, NRQCD
%\end{keywords}
\maketitle

\end{CJK*}

\section{INTRODUCTION}
\label{Introduction}

The study of heavy quarkonium has yielded valuable insight into the nature of the strong interaction, $Q\bar{Q}$ bound states have provided useful laboratories for probing both perturbative and nonperturbative QCD.
During the last years, the study of the heavy vector meson produced by photon-induced interactions at hadronic colliders has been strongly motivated by the possibility of constraining the dynamics of the strong interactions at high energies~\cite{Goncalves:2004ek,LHCForwardPhysicsWorkingGroup:2016ote,Baur:2001jj,Ma:2018zzq,Ma:2021lgv}.
It also sheds light on the low-$x$ physics and helps to constrain the nuclear parton distributions~\cite{Goncalves:2005yr,Manohar:2016nzj,Klein:2002wm,Salgado:2011wc}.
It is well known that this type of mechanism can be theoretically studied using the Weizs\"{a}cker-Williams approximation (WWA)~\cite{Fermi:1924tc,vonWeizsacker:1934nji,Phys.Rev._45_729}.
The central idea of WWA is that the moving electromagnetic field of charged particles can be treated as a flux of photons.
In an ultrarelativistic ion collider, these photons can interact with the target nucleus in the opposite beam (photoproduction) or with the photons of the opposite beam (two-photon reactions).
At the CERN Large Hadron Collider (LHC) energies, the intense heavy-ion beams represent  a prolific source of quasireal photons, hence enabling extensive studies of photon-induced physics.
In the calculations, an important function is the photon flux function, which has different forms for different charged sources.

Although great development has been achieved, the features of WWA in inelastic heavy quarkonium photoproduction in heavy-ion collisions are rarely noticed.
WWA is usually employed beyond its applicable scopes, and imprecise statements pertaining to the advantages of WWA were given~\cite{Zhu:2015via,Zhu:2015qoz,Fu:2011zzm,Fu:2011zzf,Chin.Phys.C_36_721,Yu:2015kva,Yu:2017rfi,Yu:2017pot,Drees:1989vq,Drees:1988pp,Frixione:1993yw,Nystrand:2004vn,Kniehl:2001tk,Kniehl:1990iv,Yang:2019lls,Wu:2020ujf}.
For instance, the WWA is usually adopted in electroproduction reactions or exclusive processes, where the virtuality $Q^2$ of the photon is very small, controlled by $m_e$ or the coherence condition.
However, when the WWA is used in hadronic collisions, $Q^{2}$ is controlled by the nucleus mass $m_N$ and it is not obvious that the WWA is still valid.
Particularly in the ultrarelativistic heavy-ion collisions at LHC energies, the influence of WWA becomes significant to the accuracy of describing photoproduction processes, since photon flux function scales as $f_{\gamma}\propto Z^{2}\ln\sqrt{s}/m$, in which the collision energy  $\sqrt{s}$ and the squared nuclear charge $Z^{2}$ turn into the very large enhancement factors to the cross sections.
Thus, heavy-ion collisions have a considerable flux advantage over the proton.
For these reasons, it is necessary to present a comprehensive analysis of WWA in inelastic heavy quarkonium photoproduction in heavy-ion collisions, and to estimate the important inaccuracies appeared in its application.

There are different channels which contribute to heavy quarkonium production.
From the beam side, the different photon sources need to be considered~\cite{Baur:1998ay}: coherent-photon emission (coh.), ordinary-incoherent photon emission (OIC), and ultra-incoherent photon emission (UIC).
In the first type, the virtual photons are emitted coherently by the whole nucleus which remains intact after photons radiated.
In the second and third types, the virtual photons are emitted incoherently by the protons and quarks inside nucleus, respectively, and nucleus will dissociate after photon radiation.
To avoid confusion, the terminology ``elastic'' and ``inelastic'' describe the case of whether the target nucleus remains intact or is allowed to break up after scattering with photons.
When these different photon sources are considered together, we have to weight its relative contributions to avoid double counting.
Meanwhile, in the final state, there are two types of inelastic productions need to be distinguished: direct and fragmentation contributions~\cite{Collins:1981uw,Bodwin:2015iua,Cho:1995vh}.
The fragmentation process is described by the fragmentation functions to specify the probability of final partons (gluons or quarks) hadronizing into quarkonia bound states.
The fragmentation contribution originates from the large $p_{T}$ region, where one encounters large logarithms of $p_{T}^{2}/m_{Q}^{2}$, such large logs are resummed into the fragmentation functions.
This essentially means that fragmentation mechanism can be only used in the large $p_{T}$ region, and one can not naively add the direct and fragmentation contributions together, since this will also cause double counting~\cite{Bodwin:2015iua,Bodwin:2015yma,Bodwin:2014gia}.
But in fact, the double counting exists in most works, and the fragmentation formalism is employed beyond its validity range~\cite{Zhu:2015qoz,Fu:2011zzm,Fu:2011zzf,Chin.Phys.C_36_721,Yu:2015kva, Yu:2017rfi,Yu:2017pot,Kniehl:1990iv,Drees:1988pp}.

There are numerous studies on these processes, however, the application of UIC, to our knowledge, is insufficient in inelastic heavy quarkonium photoproduction.
For instance, Gon\c{c}alves \textit{et al.} systematically studied the exclusive production of vector mesons in hadronic collisions considering different phenomenological models in Refs.~\cite{Goncalves:2019txs,Xie:2021seh}.
Machado \emph{et al.} studied the inelastic and exclusive heavy quarkonium photoproductions within the color dipole formalism~\cite{Jenkovszky:2021sis,Kopp:2018xvu}.
In Refs.~\cite{Klein:2019avl,Klein:2003vd,Baltz:2002pp}, Klein and Nystrand studied the exclusive vector meson production via photon-Pomeron or photon-meson interactions, and discussed the interlay between photoproduction and two-photon interaction~\cite{Klein:2016yzr}.
Ducati \textit{et al.} investigated the exclusive $J/\psi$ photoproduction and the radially excited $\psi(2S)$ state off nucleons in $p$-$p$ collisions according to the light-cone dipole formalism~\cite{GayDucati:2013sss}.
There are also many other relevant works, however the photon emission types in all of these works are coherent, with the incoherent-photon emission being neglected.
Furthermore, the UIC photoproduction, which is best treated as inclusive processes, can provide additional corrections to the central collisions.
For instance, the authors in Refs.~\cite{Yu:2015kva,Fu:2011zzm,Fu:2011zzf} have investigated the inelastic dileptons, photons, and light vector mesons productions at the LHC energies.
These works show that the UIC photoproduction enhances the contribution of massless and light final-state particles in the central collisions.
However, the correction is not clear for heavy quarkonium due to its large mass.

According to the above purposes, in the present work, we investigate the inelastic photoproductions of charmonium ($J/\psi$, $\psi(2S)$) and bottomonium ($\Upsilon(nS)$) in $p$-$p$ and $Pb$-$Pb$ collisions at LHC energies.
An exact treatment is performed that recovers the WWA near the $Q^{2}\sim0$ domain and can be considered as the generalization of leptoproduction~\cite{Fleming:1997fq}.
The full kinematical relations matched with the exact treatment are also obtained.
We present a consistent analysis of the features of WWA in heavy-ion collisions comparing with the exact results, and study the double counting.
In addition, we estimate the contribution of ultra-incoherent channel to the inelastic heavy quarkonia photoproductions.

It is necessary to mention that there is a important work recently done by Wangmei Zha et al., where they address similar challenges in reconciling WWA validity~\cite{Wu:2024jqn}.
Based on the QED approach, Wangmei Zha et al. developed a spatially-dependent photon flux distribution and established a more precise relationship between the photon transverse momentum distribution and impact parameter of collisions.
They justified the inadequacy of the WWA model in describing the photon flux for electron-ion collisions, and pointed out that the QED approach provides a more realistic basis for calculating the impact parameter dependence of photoproduction processes in electron-ion collisions.
Within this refined method, they explores the potential of utilizing nuetron tagging from Coulomb excitation of nuclei to effectively determine centrality for exclusive photoproduction in electron-ion collisions.
Their study offers a new methodology for exploring the spatial and momentum structure of gluons in nuclei, and offering novel insights for experimental design and data analysis.

We organize the paper as follows.
In Sec.~\ref{Exac}, we presents the formalism of exact treatment for the inelastic heavy quarkonium photoproduction in heavy-ion collisions.
Based on the Martin-Ryskin method, we consider the coh., OIC, and UIC processes simultaneously.
And according to the BCCKL method, we match the fixed order and fragmentation contributions.
In Sec.~\ref{WWA}, we turn the accurate formula into the WWA one near the region $Q^{2}\sim0$, and study the several widely utilized photon fluxes.
In Sec.~\ref{Numerical results}, we numerically calculate the $Q^{2}$, $y$, $z$, $\sqrt{s}$, and $p_{T}$ dependent differential cross sections, and the total cross sections at LHC energies.
Finally, in Sec.~\ref{Summary and conclusions} we summarize our paper.

%%%%%%%%%%%%%%%%%%%%%%%%%%%%%%%%%%%%%%%%%%%%%%%%%%%%%%%%%%%%%%%%%%%%%%%%%%%%%%%%%%%%%%%%%%%%%%%%%%%%%%%%%%%%%%
\begin{figure}
\setlength{\abovecaptionskip}{1mm}
\centering
\includegraphics[width=0.27\textwidth]{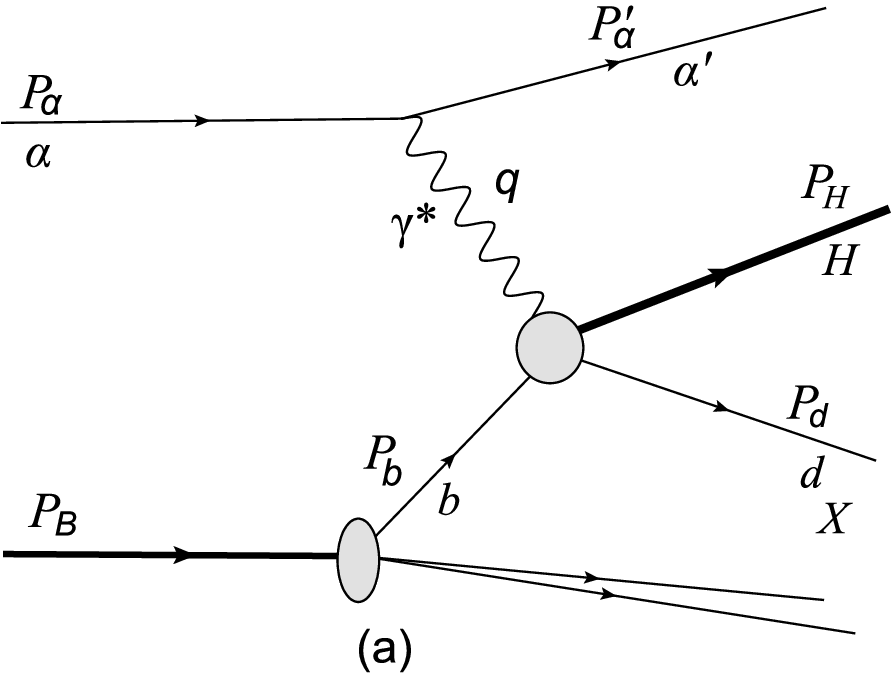}\\\hspace{60mm}
\includegraphics[width=0.20\textwidth]{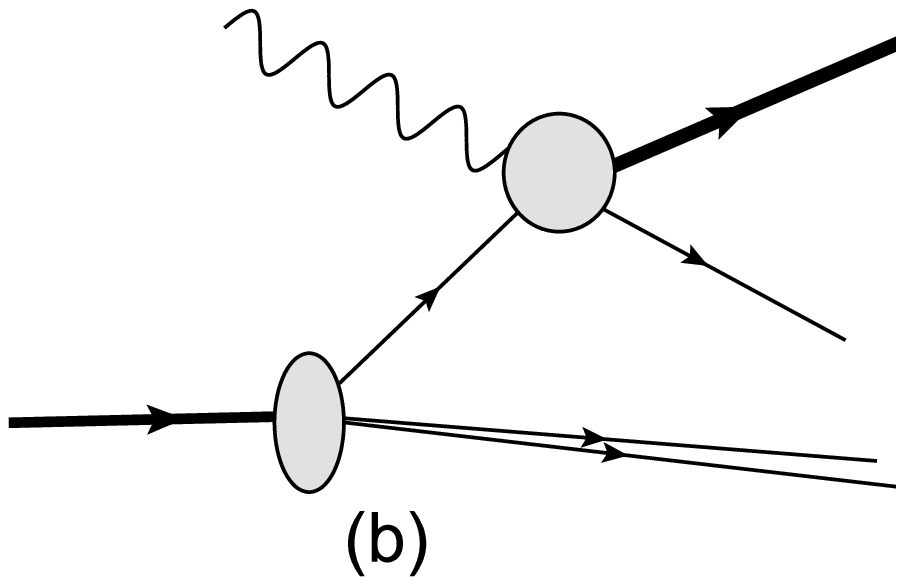}
\caption{(a): The general inelastic heavy quarkonium photoproduction.
The virtual photon emitted from $\alpha$ interacts with parton $b$ of nucleus $B$, $\alpha$ can be the nucleus or its charged parton (protons or quarks).
(b): real photoabsorption.}
\label{fig:feyn_ab}
\end{figure}
%%%%%%%%%%%%%%%%%%%%%%%%%%%%%%%%%%%%%%%%%%%%%%%%%%%%%%%%%%%%%%%%%%%%%%%%%%%%%%%%%%%%%%%%%%%%%%%%%%%%%%%%%%%%%%
\section{General formalism of Exact treatment}
\label{Exac}

For heavy quarkonium production and decay, an effective field theory known as nonrelativistic QCD (NRQCD), has been proposed to explain the huge discrepancy between the theoretical predictions and experimental measurements of the transverse momentum distribution of $J/\psi$ production at the Tevatron.
This scheme has proven to be highly successful in numerous applications~\cite{Bodwin:1994jh}.
In this section, we employ this scheme to describe the inelastic heavy quarkonium photoproduction.
The NRQCD scheme is based upon a double expansion in $\alpha_{s}$ and $\nu$ (the heavy-quark relative velocity in quarkonium rest frame), and its form for inelastic quarkonium $H$ production is

\begin{align}\label{Gen.SDCs}
d\sigma_{A+B\rightarrow H+X}=\sum_{n}d\sigma_{A+B\rightarrow Q\bar{Q}_{[1,8]}[n]+X}\langle\mathcal{O}^{H}_{[1,8]}[n]\rangle.
\end{align}
Here, $d\sigma_{A+B\rightarrow Q\bar{Q}_{[1,8]}[n]X}$ are the process-dependent short-distance coefficients (SDCs), which can be computed in perturbative QCD by expansion in $\alpha_{s}$ and correspond to the production of a heavy quark-antiquark pair  $Q\bar{Q}_{[1,8]}[n]$ in a specific color and angular momentum state $n=^{2S+1}L_{J}^{(c)}$.
The term $\langle\mathcal{O}^{H}_{[1,8]}[n]\rangle$ denotes the NRQCD long-distance matrix elements (LDMEs),  which describe the probability of hadronization and have a well-defined scaling with $v$ ($v\simeq0.08$ and $0.25$ for bottomonia and charmonia, respectively).
Since the color-singlet LDME for quarkonium production $\langle\mathcal{O}^{\psi}[^{3}S_{1}^{(1)}]\rangle$ is related to the color-singlet LDME for quarkonium decay, it can be determined in lattice QCD, from potential models, or from the $\psi$ decay rates into lepton pairs~\cite{Bodwin:2015yma}.
On the other hand, it is not known how to compute the color-octet production LDMEs from first principles, and they are usually determined by fitting some experimental data.
%The values of LDMEs may be determined by fitting some experimental data.
Notably, if NRQCD factorization holds, the LDMEs should be universal.

For $1^{--}$ quarkonia productions, $J/\psi$, $\psi\left(2S\right)$ and $\Upsilon\left(nS\right)$, the sum over $n$, truncated at order $v^{4}$, involves four LDMEs~\cite{Bodwin:1994jh}: $\langle\mathcal{O}^{H}[^{3}S_{1}^{(1)}]\rangle$, $\langle\mathcal{O}^{H}[^{3}S_{1}^{(8)}]\rangle$, $\langle\mathcal{O}^{H}[^{1}S_{0}^{(8)}]\rangle$, and $\langle\mathcal{O}^{H}[^{3}P_{J}^{(8)}]\rangle$.
Due to their lengthy expressions, we summarized them in Appendix~\ref{WF_LDMEs} for completeness and convenience.

\subsection{Accurate cross section for the general inelastic photoproduction process $\alpha b\rightarrow\alpha H X$}
\label{formalism}

The exact treatment for the inelastic heavy quarkonium photoproduction in heavy-ion collisions can be considered as proceeding in two steps.
In the first step, the density matrix of virtual photon should be expanded using the polarization operators, according to the fact that the photon radiated from the projectile is off mass shell and no longer transversely polarized.
In the second step, the square of the electric form factor $D\left(Q^{2}\right)$ is adopted as the weighting factor (WF) for different charged sources to avoid double counting.

By comprehensively analyzing the terms neglected in transiting from the exact formula of Fig.~\ref{fig:feyn_ab}(a) to the WWA one,
we can naturally estimate the features of WWA in heavy-ion collisions.
In our case, the details of the result depend essentially on the characteristic behaviour met when moving off mass shell for photoprocess amplitudes.
In the first step of exact treatment, we should derive the general form of cross section for the inelastic heavy quarkonium photoproduction in Fig.~\ref{fig:feyn_ab}(a):
\begin{align}\label{Gen.DCS}
&d\sigma\left(\alpha+B\rightarrow \alpha+H+X\right)\displaybreak[0]\nonumber\\
=&\sum_{b}\int dx_{b}f_{b/B}\left(x_{b},\mu^{2}\right)d\sigma\left(\alpha+b\rightarrow \alpha+H+d\right),\displaybreak[0]
\end{align}
where $x_{b}=p_{b}/p_{B}$ is the momentum fraction of the massless parton $b$ struck by the virtual photon, and the distribution function of parton $b$ in nucleus $B$ is
\begin{align}\label{fp}
f_{B}\left(x,\mu^{2}\right)=R_{B}\left(x,\mu^{2}\right)\left[Zp\left(x,\mu^{2}\right)+Nn\left(x,\mu^{2}\right)\right],\displaybreak[0]
\end{align}
where $R_{B}\left(x,\mu^{2}\right)$ is the nuclear modification factor~\cite{Eskola:2016oht},
$Z$ is the proton number, $N$ is the neutron number.
$p\left(x,\mu^{2}\right)$ and $n\left(x,\mu^{2}\right)$ are the parton distributions of the proton and neutron, respectively.
According to the NRQCD scheme, the partonic cross section in Eq.~(\ref{Gen.DCS}) can be written as
\begin{align}\label{Gen.pDCS}
&d\sigma\left(\alpha+b\rightarrow \alpha+H+d\right)\displaybreak[0]\nonumber\\
=&\sum_{n}\langle\mathcal{O}^{H}_{[1,8]}[n]\rangle d\sigma\left(\alpha+b\rightarrow \alpha+Q\bar{Q}_{[1,8]}[n]+d\right).\displaybreak[0]
\end{align}

Denoting the virtual photo-absorption amplitude by $M^{\mu}$, we obtain the SDCs in the parton level
\begin{align}\label{dab.Gen1}
&d\sigma\left(\alpha+b\rightarrow \alpha+Q\bar{Q}_{[1,8]}[n]+d\right)\displaybreak[0]\nonumber\\
=&\frac{4\pi e_{\alpha}^{2}\alpha_{\mathrm{em}}}{Q^2}M^{\mu}M^{*\nu}\rho_{\mu\nu}\frac{d^{3}p'_{\alpha}}{(2\pi)^{3}2E'_{\alpha}}\displaybreak[0]\nonumber\\
&\times\frac{(2\pi)^{4}\delta^{4}\left(p_{\alpha}+p_{b}-p'_{\alpha}-k\right)d\Pi}{4\sqrt{\left(p_{\alpha}\cdot p_{b}\right)^{2}-m_{\alpha}^{2}m_{b}^{2}}},\displaybreak[0]
\end{align}
where $e_{\alpha}$ is the charge of $\alpha$, $\alpha_{\mathrm{em}}$ is the electromagnetic coupling constant, $E_{\alpha'}$ is the energy of $\alpha'$,
and $\Pi$ is a phase space volume of the produced particle system (with total momentum $k$).
Keeping in mind the process in which photons may be emitted by various particles, we present a generalized density matrix of the virtual photon as:
\begin{align}\label{Rou.Gen}
\rho^{\mu\nu}=&\frac{1}{2Q^{2}}\mathrm{Tr}\left[\left(\slashed{p}_{\alpha}+m_{\alpha}\right)\Gamma^{\mu}\left(\slashed{p}'_{\alpha}+m_{\alpha}\right)\Gamma^{\nu}\right]\displaybreak[0]\nonumber\\
=&\left(-g^{\mu\nu}+\frac{q^{\mu}q^{\nu}}{q^{2}}\right)C\left(Q^{2}\right)\displaybreak[0]\nonumber\\
-&\frac{\left(2P_{\alpha}-q\right)^{\mu}\left(2P_{\alpha}-q\right)^{\nu}}{q^{2}}D\left(Q^{2}\right),\displaybreak[0]
\end{align}
where $C\left(Q^{2}\right)$ and $D\left(Q^{2}\right)$ are the general notations of form factors for $\alpha$.
Note that the $\rho^{\mu\nu}$ is non-diagonal, indicating that the virtual photons are polarized.
The expression in Eq.~(\ref{dab.Gen1}) is formulated to naturally introduce the terminology suitable for WWA.
Namely, instead of discussing nucleus-nucleus collisions [Fig.~\ref{fig:feyn_ab}(a)], one can refer to the collisions of a virtual photon off the nucleus [Fig.~\ref{fig:feyn_ab}(b)].

Now we employ the accurate expression Eq.~(\ref{dab.Gen1}) to give the $Q^{2}$- and $y$-dependent differential cross section for the inelastic heavy quarkonium photoproduction.
It is more convenient to perform the calculations in the rest frame of $\alpha$, where $|\mathbf{q}|=|\mathbf{p}_{\alpha'}|=r$, $Q^{2}=-q^{2}=\left(p_{\alpha}-p_{\alpha'}\right)^{2}=2m_{\alpha}\left(\sqrt{r^{2}+m_{\alpha}^{2}}-m_{\alpha}\right)$,
$d^{3}p_{\alpha}'=r^{2}drd\cos\theta d\varphi$, and $y=\left(q\cdot p_{b}\right)/\left(p_{\alpha}\cdot p_{b}\right)=\left(q_{0}-|\textbf{p}_{b}|r\cos\theta/E_{b}\right)/m_{\alpha}$
(which measures the relative energy loss of $\alpha$ in the lab-system).
By doing the following transformation
\begin{align}\label{Jac.Q2}
d\cos\theta dr=\mathcal{J}dQ^{2}dy=\left|\frac{D\left(r,\cos\theta\right)}{D\left(Q^{2},y\right)}\right|dQ^{2}dy,\displaybreak[0]
\end{align}
the differential cross section of Eq.~(\ref{dab.Gen1}) can be turned into (the details of $\mathcal{J}$ are given in Appendix~\ref{FKR})
\begin{align}\label{dab.Gen2}
&\frac{d\sigma\left(\alpha+b\rightarrow \alpha+Q\bar{Q}_{[1,8]}[n]+d\right)}{dQ^{2}dy}\displaybreak[0]\nonumber\\
=&\frac{e_{\alpha}^{2}\alpha_{\mathrm{em}}}{4\pi Q^2}\rho_{\mu\nu}M^{\mu}M^{*\nu}f\left(s_{\alpha b},p_{\mathrm{CM}},\hat{s},\hat{p}_{\mathrm{CM}}\right)\displaybreak[0]\nonumber\\
&\times\frac{\left(2\pi\right)^{4}\delta^{4}\left(p_{\alpha}+p_{b}-p'_{\alpha}-k\right)d\Pi}{4\hat{p}_{\mathrm{CM}}\sqrt{\hat{s}}},\displaybreak[0]
\end{align}
and
\begin{align}\label{f.kine}
&f\left(s_{\alpha b},p_{\mathrm{CM}},\hat{s},\hat{p}_{\mathrm{CM}}\right)\displaybreak[0]\nonumber\\
=&\frac{\hat{p}_{\mathrm{CM}}\sqrt{\hat{s}}}{p_{\mathrm{CM}}\sqrt{s_{\alpha b}}}
\frac{s_{\alpha b}-m_{\alpha}^{2}-m_{b}^{2}}{\sqrt{\left(s_{\alpha b}-m_{\alpha}^{2}-m_{b}^{2}\right)^{2}-4m_{\alpha}^{2}m_{b}^{2}}},\displaybreak[0]
\end{align}
where $s_{\alpha b}=(p_{\alpha}+p_{b})^{2}$ and $\hat{s}=(q+p_{b})^{2}$ are the energy square in the $\alpha$-$b$ and $\gamma^{*}$-$b$ partonic processes, respectively.
$p_{\mathrm{CM}}$ and $\hat{p}_{\mathrm{CM}}$ are the corresponding momenta.
The details are summarized in Appendix~\ref{FKR}.

After integrating over the phase space volume $\Pi$, the following quantity will be included in the result Eq.~(\ref{dab.Gen2}):
\begin{align}\label{W}
W^{\mu\nu}=\frac{1}{2}\int M^{\mu}M^{*\nu}\left(2\pi\right)^{4}\delta^{4}\left(q+p_{b}-k\right)d\Pi,\displaybreak[0]
\end{align}
where $W^{\mu\nu}$ is the absorptive part of the $\gamma b$ amplitude [Fig.~\ref{fig:feyn_ab}(b)], connected with the cross section in the usual way.
The tensors according to which $W^{\mu\nu}$ is expanded, can be constructed only from the $q$, $p_{b}$, and $g^{\mu\nu}$ tensor.
Considering the gauge invariance $q^{\mu}W^{\mu\nu}=q^{\nu}W^{\mu\nu}=0$,
it is convenient to use the following transverse and longitudinal polarization operators~\cite{Budnev:1974de}
\begin{align}\label{Polarized.Qp}
\epsilon_{T}^{\mu\nu}=&-g^{\mu\nu}+\frac{\left(q^{\mu}p_{b}^{\nu}+p_{b}^{\mu}q^{\nu}\right)}{q\cdot p_{b}}-\frac{p_{b}^{\mu}p_{b}^{\nu}q^{2}}{\left(q\cdot p_{b}\right)^{2}},\displaybreak[0]\nonumber\\
\epsilon_{L}^{\mu\nu}=&\frac{1}{q^{2}}\left(q^{\mu}-p^{\mu}\frac{q^{2}}{q\cdot p_{b}}\right)\left(q^{\nu}-p^{\nu}\frac{q^{2}}{q\cdot p_{b}}\right),\displaybreak[0]
\end{align}
which satisfy the relations: $q_{\mu}\epsilon_{T}^{\mu\nu}=q_{\mu}\epsilon_{L}^{\mu\nu}=0$, $\epsilon_{T \mu}^{\mu}=-2$, and $\epsilon_{L \mu}^{\mu}=-1$.
Furthermore,
\begin{align}\label{EpTL}
\epsilon^{\mu\nu}=\epsilon_{T}^{\mu\nu}+\epsilon_{L}^{\mu\nu}=-g^{\mu\nu}+\frac{q^{\mu}q^{\nu}}{q^{2}},\displaybreak[0]
\end{align}
is the polarization tensor of an unpolarized spin-one boson with mass $q^{2}$.
Having expanded $W^{\mu\nu}$ in these tensors, we get
\begin{align}\label{Wmn}
W^{\mu\nu}=\epsilon_{T}^{\mu\nu}W_{\mathrm{T}}\left(Q^{2},q\cdot p_{b}\right)+\epsilon_{L}^{\mu\nu}W_{\mathrm{L}}\left(Q^{2},q\cdot p_{b}\right).\displaybreak[0]
\end{align}
These Lorentz scalar functions $W_{\mathrm{T}}$ and $W_{\mathrm{L}}$ are connected with the transverse and longitudinal photon absorption cross sections $\sigma_{\mathrm{T}}$ and $\sigma_{\mathrm{L}}$, respectively:
\begin{align}\label{W.TL}
W_{\mathrm{T}}=&2\hat{p}_{\mathrm{CM}}\sqrt{\hat{s}}\sigma_{\mathrm{T}}\left(\gamma^{*}+b\rightarrow H+d\right),\displaybreak[0]\nonumber\\
W_{\mathrm{L}}=&2\hat{p}_{\mathrm{CM}}\sqrt{\hat{s}}\sigma_{\mathrm{L}}\left(\gamma^{*}+b\rightarrow H+d\right).\displaybreak[0]
\end{align}
Substituting Eqs.~(\ref{Wmn}), (\ref{W.TL}) into Eq.~(\ref{dab.Gen2}), we finally obtain
\begin{align}\label{dabTL}
&\frac{d\sigma\left(\alpha+b\rightarrow \alpha+Q\bar{Q}_{[1,8]}[n]+d\right)}{dQ^{2}dy}\displaybreak[0]\nonumber\\
=&\frac{e_{\alpha}^{2}\alpha_{\mathrm{em}}}{4\pi Q^{2}}\left[2\rho^{++}\sigma_{T}\left(\gamma^{*}+b\rightarrow Q\bar{Q}_{[1,8]}[n]+d\right)+\rho^{00}\right.\displaybreak[0]\nonumber\\
&\left.\times\sigma_{L}\left(\gamma^{*}+b\rightarrow Q\bar{Q}_{[1,8]}[n]+d\right)\right]f\left(s_{\alpha b},p_{\mathrm{CM}},\hat{s},\hat{p}_{\mathrm{CM}}\right)\displaybreak[0]\nonumber\\
=&\frac{e_{\alpha}^{2}\alpha_{\mathrm{em}}}{2\pi}d\hat{t}F_{b}[n]\left[\frac{\rho^{++}}{Q^{2}}T_{b}[n]-\rho^{00}
L_{b}[n]\right]f\left(s_{\alpha b},p_{\mathrm{CM}},\hat{s},\hat{p}_{\mathrm{CM}}\right),\displaybreak[0]\nonumber\\
\end{align}
where the relations: $d\sigma_{T}/d\hat{t}=F_{b}[n]T_{b}[n]$ and $d\sigma_{L}/d\hat{t}=-2Q^{2}F_{b}[n]L_{b}[n]$, are employed.
$F_{b}[n]$, $T_{b}[n]$ and $L_{b}[n]$ are the functions of Mandelstam variables $\hat{s}$, $\hat{t}$, $\hat{u}$, and $Q^{2}$, which can be found in Ref.~\cite{Kniehl:2001tk}.
The coefficients $\rho^{ab}$ are the elements of the density matrix Eq.~(\ref{Rou.Gen}) in the $\gamma b$-helicity basis:
\begin{align}\label{Rouzz00}
2\rho^{++}&=\epsilon_{T}^{\mu\nu}\rho_{\mu\nu}=\left[\frac{4\left(1-y\right)}{y^{2}}-\frac{4m_{\alpha}^{2}}{Q^{2}}\right]D(Q^{2})+2C\left(Q^{2}\right),\displaybreak[0]\nonumber\\
\rho^{00}&=\epsilon_{L}^{\mu\nu}\rho_{\mu\nu}=\frac{\left(2-y\right)^{2}}{y^{2}}D\left(Q^{2}\right)-C\left(Q^{2}\right).\displaybreak[0]
\end{align}

Here we come to the position to derive the second step of exact treatment,  the details of the form factors in Eq.~(\ref{Rouzz00}) need to be distinguished in each photon emission channel.
In the Martin-Ryskin method~\cite{Martin:2014nqa}, the probability or weighting factor (WF) of the coherent-photon emission is given by the square of the electric form factor in $p$-$p$ collisions: $w_{\mathrm{coh}}=G_{\mathrm{E}}^{2}\left(Q^{2}\right)=1/\left(1+Q^{2}/0.71~\mathrm{GeV}\right)^{4}$,
where the effect of the magnetic form factor is neglected.
We adopt this central idea to deal with the situation in heavy-ion collisions, where the magnetic form factor is also included.
In the case of coherent-photon emission, the photon emitter $\alpha$ is nucleus, and thus the general notations $C(Q^{2})$ and $D(Q^{2})$ in Eq.~(\ref{Rouzz00}) are the elastic nucleus form factors.
In $p$-$p$ collisions, $\alpha$ presents proton, $C(Q^{2})$ and $D\left(Q^{2}\right)$ turn into the Sachs combinations~\cite{Kniehl:1990iv}
\begin{align}\label{CDcoh.pp}
&D^{\mathrm{coh}}_{pp}\left(Q^{2}\right)=G_{\mathrm{E}}^{2}\left(Q^{2}\right)\frac{4m_{p}^{2}+7.78 Q^{2}}{4m_{p}^{2}+Q^{2}},\displaybreak[0]\nonumber\\
&C^{\mathrm{coh}}_{pp}\left(Q^{2}\right)=\mu^{2}_{p}G_{\mathrm{E}}^{2}\left(Q^{2}\right),\displaybreak[0]
\end{align}
where $\mu_{p}=2.79$ is the magnetic dipole moment.
In $Pb$-$Pb$ collisions, $\alpha$ is lead ion, $C(Q^{2})$ and $D(Q^{2})$ are
\begin{align}\label{CDcoh.PbPb}
&D^{\mathrm{coh}}_{PbPb}\left(Q^{2}\right)=Z^{2}F_{\mathrm{em}}^{2}\left(Q^{2}\right),\displaybreak[0]\nonumber\\
&C^{\mathrm{coh}}_{PbPb}\left(Q^{2}\right)=\mu^{2}_{Pb}F_{\mathrm{em}}^{2}\left(Q^{2}\right),\displaybreak[0]
\end{align}
where
\begin{align}\label{FPbem}
F_{\mathrm{em}}\left(Q^{2}\right)&=\frac{3}{(QR_{A})^{3}}\left[\sin\left(QR_{A}\right)\right.\displaybreak[0]\nonumber\\
&\left.-QR_{A}\cos\left(QR_{A}\right)\right]\frac{1}{1+a^{2}Q^{2}},\displaybreak[0]
\end{align}
is the electromagnetic form factor parameterization from the STARlight MC generator~\cite{Dyndal:2019ylt},
in which $R_{A}=1.1A^{1/3}~\mathrm{fm}$, $a=0.7~\mathrm{fm}$ and $Q=\sqrt{Q^{2}}$.

In the case of ordinary- and ultra-incoherent photon emissions, the contributions must be multiplied by the `remaining' probability, $1-w_{\mathrm{coh}}$, to avoid double counting~\cite{Martin:2014nqa}.
For ordinary-incoherent photon emission in $Pb$-$Pb$ collisions, $\alpha$ is the protons within the lead ion, and the corresponding $D(Q^{2})$ and $C(Q^{2})$ are
\begin{align}\label{CDOIC}
&D^{\mathrm{OIC}}_{PbPb}\left(Q^{2}\right)=\left[1-F_{\mathrm{em}}^{2}\left(Q^{2}\right)\right]D^{\mathrm{coh}}_{pp}\left(Q^{2}\right),\displaybreak[0]\nonumber\\
&C^{\mathrm{OIC}}_{PbPb}\left(Q^{2}\right)=\left[1-F_{\mathrm{em}}^{2}\left(Q^{2}\right)\right]C^{\mathrm{coh}}_{pp}\left(Q^{2}\right).\displaybreak[0]
\end{align}
For ultra-incoherent photon emission, $\alpha$ is the individual quarks within the nucleus.
The corresponding $C(Q^{2})$ and $D(Q^{2})$ in $p$-$p$ collisions have the following form
\begin{align}\label{CDincoh.pp}
D^{\mathrm{UIC}}_{pp}\left(Q^{2}\right)=C^{\mathrm{UIC}}_{pp}(Q^{2})=1-G_{\mathrm{E}}^{2}(Q^{2}).\displaybreak[0]
\end{align}
In $Pb$-$Pb$ collisions, since the neutron can not emit photon coherently, the WF for proton and neutron inside lead ion are different
\begin{align}\label{CDUIC}
D^{\mathrm{UIC}}_{PbPb}|_{p}(Q^{2})&=C^{\mathrm{UIC}}_{PbPb}|_{p}(Q^{2})=[1-F_{\mathrm{em}}^{2}(Q^{2})][1-G_{\mathrm{E}}^{2}(Q^{2})],\displaybreak[0]\nonumber\\
D^{\mathrm{UIC}}_{PbPb}|_{n}(Q^{2})&=C^{\mathrm{UIC}}_{PbPb}|_{n}(Q^{2})=[1-F_{\mathrm{em}}^{2}(Q^{2})].\displaybreak[0]
\end{align}

\subsection{$Q^{2}$ and $y$ distributions of heavy quarkonium production}
\label{subsec:Q2_y_dis.}

%%%%%%%%%%%%%%%%%%%%%%%%%%%%%%%%%%%%%%%%%%%%%%%%%%%%%%%%%%%%%%%%%%%%%%%%%%%%%%%%%%%%%%%%%%%%%%%%%%%%%%%%%%%%%%
\begin{figure}
\setlength{\abovecaptionskip}{1mm}
\centering
\includegraphics[width=0.205\textwidth]{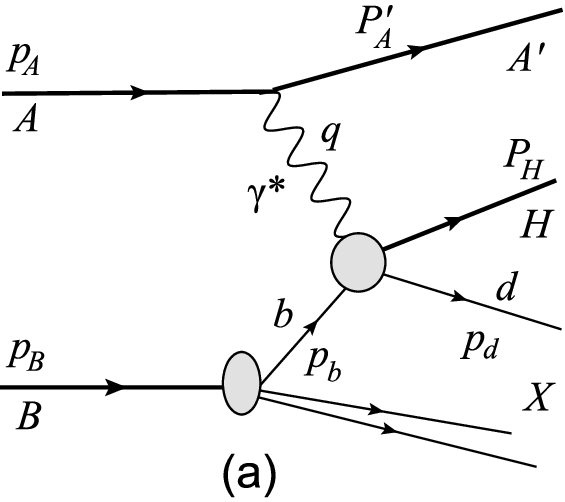}\hspace{7mm}
\includegraphics[width=0.215\textwidth]{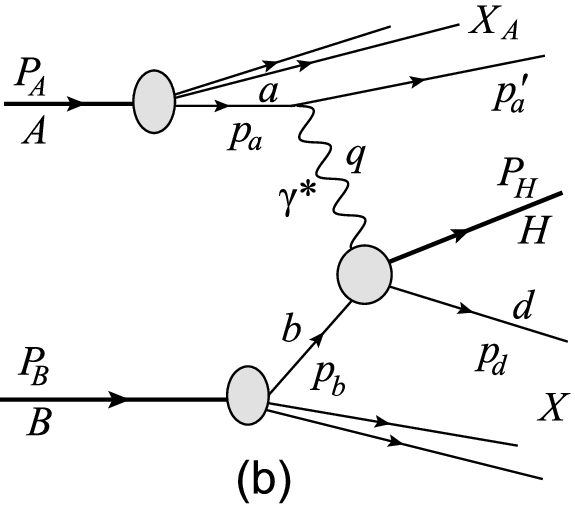}
\caption{(a): coherent-direct process in which the virtual photon emitted from the whole incident nucleus $A$ interacts with parton $b$ of target nucleus $B$ via the $\gamma^{*}$-$q$ Compton scattering and $\gamma^{*}$-$g$ fusion, and $A$ remains intact after photon emitted.
(b): incoherent-direct process in which the virtual photon emitted from the quark $a$ within nucleus $A$ interacts with parton $b$, and $A$ is allowed to break up after photon emitted.
Where photon emitter $a$ are proton and quark for ordinary-incoherent photon emission (OIC) and ultra-incoherent photon emission (UIC), respectively.
}
\label{fig:dir_pho}
\end{figure}
%%%%%%%%%%%%%%%%%%%%%%%%%%%%%%%%%%%%%%%%%%%%%%%%%%%%%%%%%%%%%%%%%%%%%%%%%%%%%%%%%%%%%%%%%%%%%%%%%%%%%%%%%%%%%%

\begin{figure}
\setlength{\abovecaptionskip}{1mm}
\centering
\includegraphics[width=0.20\textwidth]{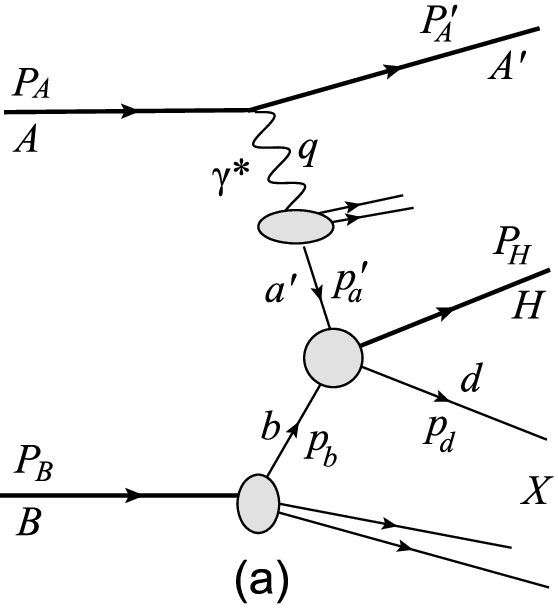}\hspace{6.6mm}
\includegraphics[width=0.22\textwidth]{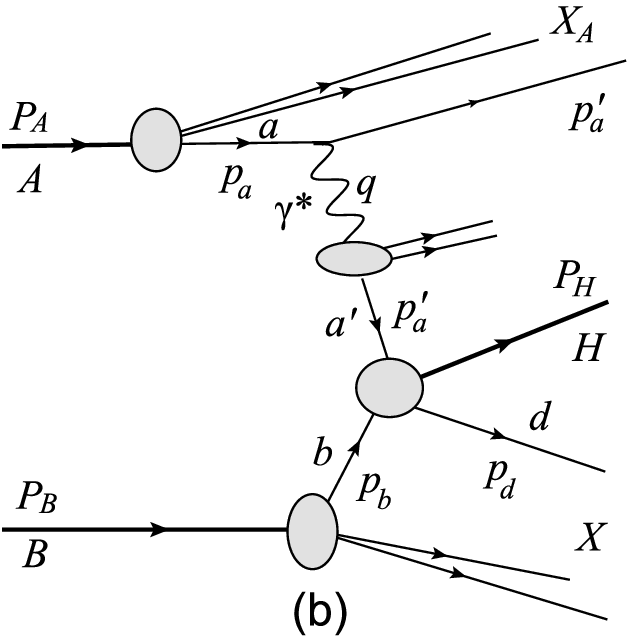}
\caption{(a): coherent-resolved process in which the parton $a'$ of hadron-like photon emitted from nucleus $A$, interacts with the parton $b$ of target $B$ via $q$-$g$ Compton scattering, $q$-$\bar{q}$ annihilation and $g$-$g$ fusion.
(b): incoherent-resolved process in which the parton $a$ inside nucleus $A$ emit a resolved virtual photon, then the parton $a^\prime$ of this resolved photon interacts with parton $b$ inside target $B$ like a hadron, and $A$ is break up after photon emitted.
Where photon emitter $a$ are proton and quark for ordinary-incoherent photon emission (OIC) and ultra-incoherent photon emission (UIC), respectively.
}
\label{fig:res_pho}
\end{figure}

Now we switch the general expression [Eq.~(\ref{dabTL})] to each specific channel in inelastic photoproduction processes in heavy-ion collisions.
In the initial state, the processes may be direct or resolved that are sensitive to the gluon distribution in the nucleus~\cite{Ma:2018zzq}.
The photons emitted from the projectile, can interact either directly with the quarks participating in the hard-scattering process (direct photoproduction) or via their quark and gluon content (resolved photoproduction).
Thus, the process $\gamma^{*}b\rightarrow H+X$ receives contributions from both direct and resolved channels.
All two contributions are formally of the same order in the perturbative expansion and must be included~\cite{Klasen:2003zn}.
Actually, as always with photons, the situation is quite complex.
Together with the three different photon emissions mentioned in Section~\ref{Introduction}, the complete description of the heavy quarkonium production requires the calculation of six classes of processes [Figs.~\ref{fig:dir_pho}-\ref{fig:res_pho}]:
coherent-direct (coh.dir.), coherent-resolved (coh.res.), ordinary-incoherent direct (OIC dir.), ordinary-incoherent resolved (OIC res.), ultra-incoherent direct (UIC dir.), and ultra-incoherent resolved (UIC res.) processes.
These abbreviations will appear in many places of the remaining content.

%%%%%%%%%%%%%%%%%%%%%%%%%%%%%%%%%%%%%%%%%%%%%%%%%%%%%%%%%%%%%%%%%%%%%%%%%%%%%%%%%%%%%%%%%%%%%%%%%%%%%%%%%%%%%%
%%%%%%%%%%%%%%%%%%%%%%%%%%%%%%%%%%%%%%%%%%%%%%%%%%%%%%%%%%%%%%%%%%%%%%%%%%%%%%%%%%%%%%%%%%%%%%%%%%%%%%%%%%%%%%

In the case of direct photoproduction [Fig.~\ref{fig:dir_pho}], the corresponding differential cross sections are
\begin{align}\label{CSyQ2.dir}
&\frac{d\sigma_{\mathrm{coh.dir.}}}{dQ^{2}dy}\left(A+B\rightarrow A+H+X\right)\displaybreak[0]\nonumber\\
=&2\sum_{b}\int dx_{b}d\hat{t}f_{b/B}\left(x_{b},\mu^{2}\right)\sum_{n}\langle\mathcal{O}^{H}_{[1,8]}[n]\rangle\displaybreak[0]\nonumber\\
&\times\frac{d\sigma}{dydQ^{2}d\hat{t}}\left(A+b\rightarrow A+Q\bar{Q}_{[1,8]}[n]+d\right),\displaybreak[0]\\
\displaybreak[0]\nonumber\\
&\frac{d\sigma_{\mathrm{OIC~dir.}}}{dQ^{2}dy}\left(A+B\rightarrow X_{A}+H+X\right)\displaybreak[0]\nonumber\\
=&2Z_{Pb}\sum_{b}\int dx_{b}d\hat{t}f_{b/B}\left(x_{b},\mu^{2}\right)\sum_{n}\langle\mathcal{O}^{H}_{[1,8]}[n]\rangle\displaybreak[0]\nonumber\\
&\times\frac{d\sigma}{dydQ^{2}d\hat{t}}\left(p+b\rightarrow p+Q\bar{Q}_{[1,8]}[n]+d\right).\displaybreak[0]\\
\displaybreak[0]\nonumber\\
&\frac{d\sigma_{\mathrm{UIC~dir.}}}{dQ^{2}dy}\left(A+B\rightarrow X_{A}+H+X\right)\displaybreak[0]\nonumber\\
=&2\sum_{a,b}\int dx_{a}dx_{b}d\hat{t}f_{a/A}\left(x_{a},\mu^{2}\right)f_{b/B}\left(x_{b},\mu^{2}\right)\sum_{n}\langle\mathcal{O}^{H}_{[1,8]}[n]\rangle\displaybreak[0]\nonumber\\
&\times\frac{d\sigma}{dydQ^{2}d\hat{t}}\left(a+b\rightarrow a+Q\bar{Q}_{[1,8]}[n]+d\right),\displaybreak[0]
\end{align}
where the factor of two arises because both nuclei emit photons and thus serve as targets.
The partonic cross section can be derived from Eq.~(\ref{dabTL}) with $m_{\alpha}=m_{q}=0$ and $e_{\alpha}=e_{a}$, where $e_{a}$ is the charge of massless quark $a$.

In the resolved photoproduction [Fig.~\ref{fig:res_pho}], the corresponding differential cross sections are
\begin{align}\label{CSyQ2.res}
&\frac{d\sigma_{\mathrm{coh.res.}}}{dQ^{2}dy}\left(A+B\rightarrow A+H+X\right)\displaybreak[0]\nonumber\\
=&2\sum_{b}\sum_{a'}\int dx_{b}dz_{a'}d\hat{t}f_{b/B}\left(x_{b},\mu^{2}\right)f_{a'/\gamma}\left(z_{a'},\mu^{2}\right)\displaybreak[0]\nonumber\\
&\times\frac{Z_{Pb}^{2}\alpha_{\mathrm{em}}}{2\pi}\frac{y\rho^{++}_{\mathrm{coh}}}{Q^{2}}\sum_{n}\langle\mathcal{O}^{H}_{[1,8]}[n]\rangle\frac{d\sigma_{a'b\rightarrow Q\bar{Q}_{[1,8]}[n]d}}{d\hat{t}},\displaybreak[0]\\
\displaybreak[0]\nonumber\\
&\frac{d\sigma_{\mathrm{OIC~res.}}}{dQ^{2}dy}\left(A+B\rightarrow X_{A}+H+X\right)\displaybreak[0]\nonumber\\
=&2Z_{Pb}\sum_{b}\sum_{a'}\int dx_{b}dz_{a'}d\hat{t}f_{b/B}\left(x_{b},\mu^{2}\right)f_{a'/\gamma}\left(z_{a'},\mu^{2}\right)\displaybreak[0]\nonumber\\
&\times \frac{\alpha_{\mathrm{em}}}{2\pi}\frac{y\rho^{++}_{\mathrm{OIC}}}{Q^{2}}
\sum_{n}\langle\mathcal{O}^{H}_{[1,8]}[n]\rangle\frac{d\sigma_{a'b\rightarrow Q\bar{Q}_{[1,8]}[n]d}}{d\hat{t}}.\displaybreak[0]\\
\displaybreak[0]\nonumber\\
&\frac{d\sigma_{\mathrm{UIC~res.}}}{dQ^{2}dy}\left(A+B\rightarrow X_{A}+H+X\right)\displaybreak[0]\nonumber\\
=&2\sum_{a,b}\sum_{a'}\int dx_{a}dx_{b}dz_{a'}d\hat{t}f_{a/A}\left(x_{a},\mu^{2}\right)\displaybreak[0]\nonumber\\
&\times f_{b/B}\left(x_{b},\mu^{2}\right)f_{a'/\gamma}\left(z_{a'},\mu^{2}\right)\frac{e_{a}^{2}\alpha_{\mathrm{em}}}{2\pi}\frac{y\rho^{++}_{\mathrm{UIC}}}{Q^{2}}\displaybreak[0]\nonumber\\
&\times \sum_{n}\langle\mathcal{O}^{H}_{[1,8]}[n]\rangle\frac{d\sigma_{a'b\rightarrow Q\bar{Q}_{[1,8]}[n]d}}{d\hat{t}}.\displaybreak[0]
\end{align}
where $z_{a}'=p_{a'}/q$ and $f_{\gamma}\left(z_{a'},\mu^{2}\right)$ denote the parton's momentum fraction and the parton distribution function of the resolved photon~\cite{Gluck:1999ub}, respectively.
The involved partonic cross sections can be found in Ref.~\cite{Klasen:2003zn}.

\subsection{$p_{T}$ and $z$ distributions of heavy quarkonium production}
\label{pT distribution}

The distributions in $p_{T}$ and inelastic variable $z=(p_{H}\cdot p_{b})/(q\cdot p_{b})$ can be obtained using the Jacobian transformation.
In the final state, we need to classify the two types of inelastic photoproductions.
In the first type, direct heavy quarkonium produced from the $\gamma$-$g$ fusion, annihilation and Compton scattering of partons.
In the second type, fragmentation heavy quarkonium produced through the final fragmentation of a parton.
In the following, we will take into account all of these aspects.

\subsubsection{Direct heavy quarkonium photoproduction}
\label{sec:D_HQ_PT_yr_z}

Before doing the transformation the Mandelstam variables in $\gamma^{*}$-$b$ parton level should be expressed as:
\begin{align}\label{Mant.yr}
\hat{s}=&\left(M_{T}\cosh y_{r}+\sqrt{\cosh^{2}y_{r}M_{T}^{2}+m_{b}^{2}-M_{H}^{2}}\right)^{2},\displaybreak[0]\nonumber\\
\hat{t}=&M_{H}^{2}-Q^{2}-2M_{T}\left(\hat{E}_{\gamma}\cosh y_{r}-\hat{p}_{\mathrm{CM}}\sinh y_{r}\right),\displaybreak[0]\nonumber\\
\hat{u}=&M_{H}^{2}-2M_{T}\left(\hat{E}_{b}\cosh y_{r}+\hat{p}_{\mathrm{CM}}\sinh y_{r}\right),\displaybreak[0]
\end{align}
where $y_{r}$ is the rapidity, $M_{T}=\sqrt{M_{H}^{2}+p_{T}^{2}}$ is the transverse mass of heavy quarkonium, $\hat{E}_{\gamma}$, $\hat{E}_{b}$, and $\hat{p}_{\mathrm{CM}}$ are the corresponding energies and momentum.
The details are summarized in Appendix~\ref{FKR}.

In the case of direct-photon processes, the variables $x_{b}$ and $\hat{t}$ should be transformed as
\begin{align}\label{Jac.dir.}
d\hat{t}dx_{b}=\mathcal{J}dy_{r}dp_{T}=\left|\frac{D\left(x_{b},\hat{t}\right)}{D\left(y_{r},p_{T}\right)}\right|dy_{r}dp_{T},\displaybreak[0]
\end{align}
and the corresponding differential cross sections are
\begin{align}\label{dPT.dir.}
&\frac{d\sigma_{\mathrm{coh.dir.}}}{dp_{T}dy_{r}}\left(A+B\rightarrow A+H+X\right)\displaybreak[0]\nonumber\\
=&2\sum_{b}\int dQ^{2}dyf_{b/B}\left(x_{b},\mu^{2}\right)\mathcal{J}\sum_{n}\langle\mathcal{O}^{H}_{[1,8]}[n]\rangle\nonumber\\
&\times\frac{d\sigma}{dQ^{2}dyd\hat{t}}\left(A+b\rightarrow A+Q\bar{Q}_{[1,8]}[n]+d\right),\displaybreak[0]\\
&\displaybreak[0]\nonumber\\
&\frac{d\sigma_{\mathrm{OIC~dir.}}}{dp_{T}dy_{r}}\left(A+B\rightarrow X_{A}+H+X\right)\displaybreak[0]\nonumber\\
=&2Z_{Pb}\sum_{b}\int dQ^{2}dyf_{b/B}\left(x_{b},\mu^{2}\right)\mathcal{J}\sum_{n}\langle\mathcal{O}^{H}_{[1,8]}[n]\rangle\nonumber\\
&\times\frac{d\sigma}{dQ^{2}dyd\hat{t}}\left(p+b\rightarrow p+Q\bar{Q}_{[1,8]}[n]+d\right),\displaybreak[0]\\
\displaybreak[0]\nonumber\\
&\frac{d\sigma_{\mathrm{UIC~dir.}}}{dp_{T}dy_{r}}\left(A+B\rightarrow X_{A}+H+X\right)\displaybreak[0]\nonumber\\
=&2\sum_{a,b}\int dQ^{2}dydx_{a}f_{a/A}\left(x_{a},\mu^{2}\right)f_{b/B}\left(x_{b},\mu^{2}\right)\mathcal{J}\nonumber\\
&\times\sum_{n}\langle\mathcal{O}^{H}_{[1,8]}[n]\rangle\frac{d\sigma}{dQ^{2}dyd\hat{t}}\left(a+b\rightarrow a+Q\bar{Q}_{[1,8]}[n]+d\right).\displaybreak[0]
\end{align}

In the case of resolved processes, we should choose the variables $\hat{t}^{*}$ and $z_{a'}$ to do the similar transformation
\begin{align}\label{Jac.res.}
d\hat{t}^{*}dz_{a'}=\mathcal{J}dy_{r}dp_{T}=
\left|\frac{D\left(z_{a'},\hat{t}^{*}\right)}{D\left(y_{r},p_{T}\right)}\right|dy_{r}dp_{T},\displaybreak[0]
\end{align}
the corresponding differential cross sections are
\begin{align}\label{dPT.res.}
&\frac{d\sigma_{\mathrm{coh.res.}}}{dp_{T}dy_{r}}\left(A+B\rightarrow A+H+X\right)\displaybreak[0]\nonumber\\
=&2\sum_{b}\sum_{a'}\int dQ^{2}dydx_{b}f_{b/B}\left(x_{b},\mu^{2}\right)f_{\gamma}\left(z_{a'},\mu^{2}\right)\mathcal{J}\displaybreak[0]\nonumber\\
&\times \frac{Z_{Pb}^{2}\alpha_{\mathrm{em}}}{2\pi}\frac{y\rho^{++}_{\mathrm{coh}}}{Q^{2}}
\sum_{n}\langle\mathcal{O}^{H}_{[1,8]}[n]\rangle\frac{d\sigma_{a'b\rightarrow Q\bar{Q}_{[1,8]}[n]d}}{d\hat{t}},\displaybreak[0]\\
&\displaybreak[0]\nonumber\\
&\frac{d\sigma_{\mathrm{OIC res.}}}{dp_{T}dy_{r}}\left(A+B\rightarrow X_{A}+H+X\right)\displaybreak[0]\nonumber\\
=&2Z_{Pb}\sum_{b}\sum_{a'}\int dQ^{2}dydx_{b}f_{b/B}\left(x_{b},\mu^{2}\right)f_{\gamma}\left(z_{a'},\mu^{2}\right)\displaybreak[0]\nonumber\\
&\times \mathcal{J}\frac{\alpha_{\mathrm{em}}}{2\pi}\frac{y\rho^{++}_{\mathrm{OIC}}}{Q^{2}}
\sum_{n}\langle\mathcal{O}^{H}_{[1,8]}[n]\rangle\frac{d\sigma_{a'b\rightarrow Q\bar{Q}_{[1,8]}[n]d}}{d\hat{t}},\displaybreak[0]\\
&\displaybreak[0]\nonumber\\
&\frac{d\sigma_{\mathrm{UIC res.}}}{dp_{T}dy_{r}}\left(A+B\rightarrow X_{A}+H+X\right)\displaybreak[0]\nonumber\\
=&2\sum_{a,b}\sum_{a'}\int dQ^{2}dydx_{a}dx_{b}f_{a/A}\left(x_{a},\mu^{2}\right)\displaybreak[0]\nonumber\\
&\times f_{b/B}\left(x_{b},\mu^{2}\right)f_{\gamma}\left(z_{a'},\mu^{2}\right)\mathcal{J}e_{a}^{2}\frac{\alpha_{\mathrm{em}}}{2\pi}\frac{y\rho^{++}_{\mathrm{UIC}}}{Q^{2}}\displaybreak[0]\nonumber\\
&\times\sum_{n}\langle\mathcal{O}^{H}_{[1,8]}[n]\rangle\frac{d\sigma_{a'b\rightarrow Q\bar{Q}_{[1,8]}[n]d}}{d\hat{t}},\displaybreak[0]\label{dPT.Us.}
\end{align}
where the Mandelstam variables of resolved photoproductions are the same as Eq.~(\ref{Mant.yr}) but for $Q^{2}=0$.

For the $z$ distribution, we should rewrite the Mandelstam variables in Eq.~(\ref{Mant.yr}) as follows:
\begin{align}\label{Mant.z}
&\hat{s}=\frac{M_{H}^{2}}{z}+\frac{p_{T}^{2}}{z\left(1-z\right)},\displaybreak[0]\nonumber\\
&\hat{t}=-\left(1-z\right)\left(\hat{s}+Q^{2}\right),\displaybreak[0]\nonumber\\
&\hat{u}=M_{H}^{2}-z\left(\hat{s}+Q^{2}\right),\displaybreak[0]
\end{align}
thus, we can do the similar Jacobian transformation.
The relevant cross sections in $z$ distribution can be obtained in the following manner
\begin{align}\label{Mant.z}
\frac{d\sigma}{dz}=\int^{p^{2}_{T\mathrm{max}}}_{p^{2}_{T\mathrm{min}}}\frac{d^{2}\sigma}{dzdp^{2}_{T}}dp^{2}_{T}.\displaybreak[0]
\end{align}
The Jacobian factors $\mathcal{J}$ and corresponding kinematical boundaries are summarized in Appendix~\ref{FKR}.

\subsubsection{Fragmentation heavy quarkonium production}
%%%%%%%%%%%%%%%%%%%%%%%%%%%%%%%%%%%%%%%%%%%%%%%%%%%%%%%%%%%%%%%%%%%%%%%%%%%%%%%%%%%%%%%%%%%%%%%%%%%%%%%%%%%%%%
\label{sec:F_HQ_PT_yr_z}
\begin{figure}
\setlength{\abovecaptionskip}{1mm}
\centering
\includegraphics[width=0.48\textwidth]{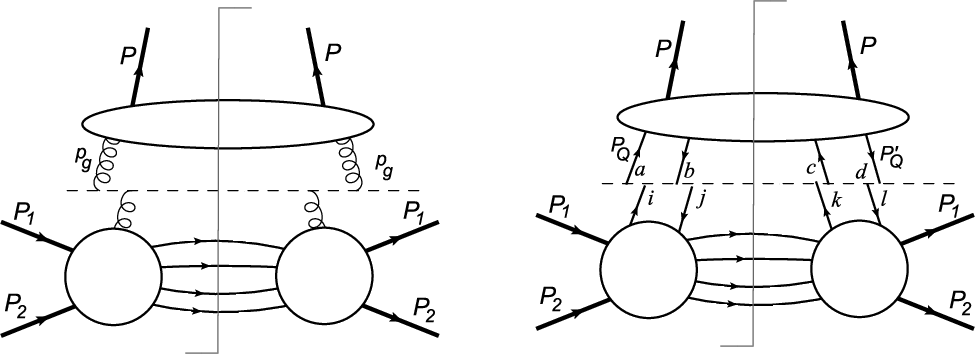}\hspace{7mm}
\caption{pQCD factorization diagrams of heavy quarkonium production. Left: sing-parton (here taking gluon as an example) fragmentation.
  Right: heavy quark-pair fragmentation.}
\label{fig:LP_NLP}
\end{figure}
%%%%%%%%%%%%%%%%%%%%%%%%%%%%%%%%%%%%%%%%%%%%%%%%%%%%%%%%%%%%%%%%%%%%%%%%%%%%%%%%%%%%%%%%%%%%%%%%%%%%%%%%%%%%%%
The fragmentation heavy quarkonium production is also an important channel which is described by the fragmentation functions (FFs).
This factorization theory of quarkonium production that has been proven at present is the collinear factorization method~\cite{Kang:2011mg}, where one can compute rates at leading power (LP) or next to leading power (NLP) in $m_{Q}^{2}/p_{T}^{2}$.
LP contributions can be factorized into partonic cross sections to produce a specific single parton convolved with one particle FFs~\cite{Collins:1981uw}.
NLP contributions can be factorized to produce two specific partons convolved with two-parton FFs~\cite{Kang:2011mg}.
The Feynman diagrams in the cut diagram notation for these two contributions are shown in Fig.~\ref{fig:LP_NLP}.
This fragmentation picture dominates over all other creation mechanisms at large $p_{T}$ range.
In particular, gluon fragmentation represents the dominant source of high energy prompt quarkonia at hadron colliders.
Of course, because the LP and NLP contributions represent the leading and first subleading terms in an expansion in powers of $m_{Q}^{2}/p_{T}^{2}$, one would not expect them to be valid unless $p_{T}$ is significantly greater than $m_{Q}$.
%the approximations that enter into FFs computations break down when a quarkonium's energy becomes comparable to its mass.
Fragmentation predictions for charmonium differential cross sections are therefore unreliable at low $p_{T}$ domain.
In order to suppress NLP contributions and to ensure the validity of fragmentation mechanism, Ref.~\cite{Faccioli:2014cqa} suggested that a criterion $p_{T}>3M_{J/\psi}$ be used in comparing data with theory.
In Refs.~\cite{Bodwin:2015iua,Bodwin:2015yma,Bodwin:2014gia}, a criterion $p_{T}>10~\textrm{GeV}$ was used in their predictions.
Similarly, fragmentation results for bottomonium production are untrustworthy throughout the $p_{T}<15~\mathrm{GeV}$ region~\cite{Cho:1995vh}.

In present paper, we adopt the LP-factorization formalism to calculate the fragmentation contributions to heavy quarkonium production.
Calculations of these fragmentation contributions, at any given order in $\alpha_{s}$, are much simpler than those of full fixed-order calculations.
According to the NRQCD scheme, the inelastic quarkonium $H$ production in two-body collisions can be written as Eq.~(\ref{Gen.SDCs}), where the SDCs, $d\sigma_{A+B\rightarrow Q\bar{Q}_{[1,8]}[n]X}$, describes the production of a $Q\bar{Q}$ pair in color and angular momentum state $n$.
At large transverse momentum $p_{T}$, one can apply LP factorization to the SDCs in Eq.~(\ref{Gen.SDCs}).
The general expression is~\cite{Collins:1981uw, Nayak:2005rt}
\begin{align}\label{LP.Gen.}
d\sigma^{\textrm{LP-frag}}_{A+B\rightarrow Q\bar{Q}[n]+X}
=\sum_{c}d\hat{\sigma}_{A+B\rightarrow c+X}\otimes D_{c\rightarrow Q\bar{Q}[n]}.\displaybreak[0]
\end{align}
where $d\hat{\sigma}_{A+B\rightarrow c+X}$ are the parton production cross sections (PPCSs) to produce a parton $c$, and $D_{c\rightarrow Q\bar{Q}[n]}$ are FFs for a parton $c$ to fragment into a $Q\bar{Q}$ pair with quantum numbers $n$.
In the LP factorization, the SDCs of Eq.~(\ref{LP.Gen.}) for photoproduction and initial partons hard scattering (had.scat.) can be summarized as following master formula
\begin{align}\label{LP.Frag.spe}
%&d\sigma_{\mathrm{had.scat.}}\displaybreak[0]\nonumber\\
%=&\sum_{a,b}\int dx_{a}dx_{b}f_{a/A}(x_{a},\mu^{2})f_{b/B}(x_{b},\mu^{2})\displaybreak[0]\nonumber\\
%&\times\sum_{n}\langle\mathcal{O}^{H}[n]\rangle d\sigma_{ab\rightarrow Q\bar{Q}_{[1,8]}[n]d},\displaybreak[0]\\
%\displaybreak[0]\nonumber\\
&d\sigma^{\textrm{LP-frag}}_{A+B\rightarrow Q\bar{Q}_{[1,8]}[n]X}\displaybreak[0]\nonumber\\
=&\sum_{a,b,c}\sum_{a'}f_{a/A}\left(x_{a},\mu^{2}\right)\otimes f_{k/a}\left(z_{a'},\mu^{2}\right)\otimes f_{b/B}\left(x_{b},\mu^{2}\right)\otimes\displaybreak[0]\nonumber\\
&d\sigma_{kb\rightarrow cX}(p_{c}=p'_{c}/z_{c},\mu^{2})\otimes D_{c\rightarrow Q\bar{Q}_{[1,8]}[n]}\left(z_{c},\mu^{2}\right),\displaybreak[0]
\end{align}
where $z_{c}$ is the longitudinal momentum fraction of the hadron relative to the fragmenting parton.
$f_{a/A}\left(x_{a},\mu^{2}\right)$ is the PDF of parton $a=g,q,\bar{q}$ in nucleus A, or the flux function of photon $a=\gamma$.
$f_{k/a}\left(z_{a'},\mu^{2}\right)$ is $\delta_{ak}\delta\left(1-z_{a'}\right)$ in the ``direct'' case (since the parton $a$ is the photon itself), or the PDF of parton $k$ in the resolved photon $a$.
The parton distributions in the photon behave like $\alpha/\alpha_{s}(Q^2)$ for large $Q^2$.
Therefore the additional power of $\alpha_{s}$ contained in the ``resolved'' component as compared to the ``direct'' one is compensated.
The hard PPCSs, $d\sigma_{kb\rightarrow cX}$, and FFs have been calculated to NLO accuracy, they are expansions in powers of $\alpha_{s}$
\begin{align}\label{PPCSs}
d\hat{\sigma}_{\gamma b\rightarrow cX}&=\alpha_{s}d\hat{\sigma}^{(1)}_{\gamma b\rightarrow cX}+\alpha_{s}^{2}d\hat{\sigma}^{(2)}_{\gamma b\rightarrow cX}+\mathcal{O}(\alpha_{s}^{3}),\displaybreak[0]\nonumber\\
d\hat{\sigma}_{ab\rightarrow cX}&=\alpha_{s}^{2}d\hat{\sigma}^{(2)}_{\gamma b\rightarrow cX}+\alpha_{s}^{3}d\hat{\sigma}^{(3)}_{\gamma b\rightarrow cX}+\mathcal{O}(\alpha_{s}^{4}),\displaybreak[0]\nonumber\\
D_{c\rightarrow Q\bar{Q}[n]}&=\alpha_{s}D^{(1)}_{c\rightarrow Q\bar{Q}[n]}+\alpha_{s}^{2}D^{(2)}_{c\rightarrow Q\bar{Q}[n]}+\mathcal{O}(\alpha_{s}^{3}).\displaybreak[0]
\end{align}

In this paper, we use the LP factorization approximation~\label{LP.Frag.} for the SDCs to compute fragmentation contributions that augment the LO fixed order calculations of the SDCs which are presented in the previous sections.
Since direct and fragmentation contributions can not be directly added up, we adopt the method developed by Bodwin, Chao, Chung, Kim, and Lee (BCCKL) to avoid double counting, where the matching rule between fixd-order and fragmentation has been performed~\cite{Bodwin:2015iua,Bodwin:2015yma,Bodwin:2014gia}.
Therefore, we combine the LO fixed-order calculations with LP fragmentation corrections [Eq.~(\ref{LP.Gen.})] according to the formula
\begin{align}\label{LO.LP.}
d\sigma=d\sigma_{\textrm{LO}}(\alpha_{s}^{3})+d\sigma^{\textrm{LP-frag.}}(\alpha_{s}^{4}),\displaybreak[0]
\end{align}
where we denotes the LO ($\alpha_{s}^{3}$) contributions to SDCs by $d\sigma_{\textrm{LO}}$, which are discussed in previous sections and start from Eq.~(\ref{dab.Gen1}).
We denotes the LP fragmentation corrections by $d\sigma^{\textrm{LP-frag.}}$, and only consider the contributions at order $\alpha_{s}^{4}$, because the authors in Ref.~\cite{Bodwin:2015yma} have shown that the LP fragmentation contributions at order $\alpha_{s}^{5}$ are small, and only have little effect on the cross section prediction for photoproduction.
%A summary of FFs that we used here is given in Refs.~\cite{Braaten:1994kd,Ma:2013yla}

For $d\sigma^{\textrm{LP-frag.}}$, we combine the PPCSs and FFs both at order $\alpha_{s}^{2}$.
The LP fragmentation process also include the six classes of sub-processes which are described in Figs.~\ref{fig:dir_pho}, \ref{fig:res_pho}.
According to Eq.~(\ref{LP.Gen.}), in calculations we just need to replace the LO SDCs [Eqs.~(\ref{dPT.dir.})-(\ref{dPT.Us.})] with the convolution of PPCSs and FFs.
%There are two types of photon-induced process that contribute to photoproduction cross sections: direct and resolved processes.
The PPCSs of the direct and resolved processes can be found in Refs.~\cite{Ma:2019mwr,Owens:1986mp}.
We consider gluon and light-quark fragmentations.
The gluon FFs $D_{g\rightarrow Q\bar{Q}[n]}$ are given for the $^{1}S_{0}^{(8)}$ at order $\alpha_{s}^{2}~(\textrm{LO})$ in Refs.~\cite{Braaten:1996rp,Bodwin:2012xc}, for the $^{3}S_{1}^{(8)}$ at orders $\alpha_{s}~(\textrm{LO})$ and $\alpha_{s}^{2}~(\textrm{NLO})$ in Refs.~\cite{Braaten:1994kd,Ma:2013yla}, and for the $^{3}P_{J}^{(8)}$ at order $\alpha_{s}^{2}~(\textrm{LO})$ in Refs.~\cite{Bodwin:2012xc,Braaten:1994kd}.
Because the gluon FF for the $^{3}S_{1}^{(1)}$ begins at order $\alpha_{s}^{3}~(\textrm{LO})$,
%thus the $^{3}S_{1}^{(1)}$ channel receives an LP contribution that begins at $\alpha_{s}^{5}$~\cite{Bodwin:2015yma,Bodwin:2015iua}.
we do not consider it here.
%And since the $^{3}S_{1}^{(8)}$ contribution is much smaller than those of $^{1}S_{0}^{(8)}$ and $^{3}P_{J}^{(8)}$, we also neglected its contribution.
The light quark FF $D_{q\rightarrow Q\bar{Q}}[n]$ is given for the $^{3}S_{1}^{(8)}$ at order $\alpha_{s}^{2}~(\textrm{LO})$ in Refs.~\cite{Ma:2013yla,Ma:1995vi}, the other light quark FFs vanishes though order $\alpha_{s}^{2}$.
%And we ignore the color-singlet LP fragmentation contribution ($^{3}S_{1}^{(1)}$) in our numerical results.
%Since at very large $p_{T}$ domain, color octet dominates the singlet contribution~\cite{Ko:1996xw}.

%the involved partonic processes for had.scat. are $gg\rightarrow Q\bar{Q}_{[1,8]}[n]g$, $gq\rightarrow Q\bar{Q}_{[1,8]}[n]q$, and $q\bar{q}\rightarrow Q\bar{Q}_{[1,8]}[n]g$~\cite{Owens:1986mp}.
%Firstly, we should rewrite the Mandelstam variables as the following forms

For achieving $p_{T}$ distribution, we should rewrite the Mandelstam variables as follows,
\begin{align}\label{Mant.frag.}
&\hat{s}=y\left(s_{\alpha b}-m_{\alpha}^{2}-m_{b}^{2}\right)-Q^{2}+m_{b}^{2},\displaybreak[0]\nonumber\\
&\hat{t}=\frac{1}{2\cosh y_{r}}\left[Q^{2}\left(e^{y_{r}}-2\cosh y_{r}\right)-e^{-y_{r}}\hat{s}\right],\displaybreak[0]\nonumber\\
&\hat{u}=-\left(\hat{s}+Q^{2}\right)\frac{e^{y_{r}}}{2\cosh y_{r}},\displaybreak[0]
\end{align}
and since the parton $c$ is taken to be lightlike by neglecting the parton mass, $z_{c}=p'_{c}/p_{c}=2p_{T}\cosh y_{r}/\sqrt{\hat{s}}$.
Then the variables $z_{c}$ and $\hat{t}$ can do the transformation
\begin{align}\label{Jac.frag}
d\hat{t}dz_{c}=\mathcal{J}dy_{r}dp_{T}=\left|\frac{D\left(z_{c},\hat{t}\right)}{D\left(y_{r},p_{T}\right)}\right|dy_{r}dp_{T}.\displaybreak[0]
\end{align}

\section{Weizs\"{a}cker-Williams approximation}
\label{WWA}
The connection between the process in Fig.~\ref{fig:feyn_ab}(a) and (b) is evident.
By Fourier-transforming the electric and magnetic fields of an ultrarelativistic charged point-like particle, the photoproduction process can be expressed in terms of the real photo-absorption cross section with the photon spectrum.
This idea was originally pointed out by Fermi~\cite{Fermi:1924tc}, and was independently developed for the process involving relativistic collisions of charged particles by Weizs\"{a}cker and Williams, and the method is now known as the Weizs\"{a}cker-Williams approximation (WWA)~\cite{vonWeizsacker:1934nji}.
An essential advantage of WWA consists in the fact that, when using it, it is sufficient to obtain the photo-absorption cross section on the mass shell only.
Details of its off mass-shell behavior are not essential.
In the present section, we switch the accurate expression Eq.~(\ref{dabTL}) into the WWA form by taking $Q^{2}\rightarrow0$, and discuss a number of widely employed photon spectra.
There are two simplifications: the scalar photon contribution $\sigma_{L}$ is neglected;
the term of $\sigma_{T}$ is substituted by its on-shell value.
This provides us a powerful approach to study the features of WWA in inelastic heavy quarkonium photoproduction in heavy-ion collisions.

Taking $Q^{2}\rightarrow0$, Eq.~(\ref{dabTL}) turns into:
\begin{align}\label{dWWA.Gen.}
&\lim_{Q^{2}\rightarrow0}d\sigma\left(\alpha+b\rightarrow\alpha+Q\bar{Q}_{[1,8]}+d\right)\displaybreak[0]\nonumber\\
=&\left[\frac{e_{\alpha}^{2}\alpha_{\mathrm{em}}}{2\pi}(y\rho^{++})\frac{dydQ^{2}}{Q^{2}}\right]
\sigma_{T}\frac{\hat{p}_{\mathrm{CM}}\sqrt{\hat{s}}}{yp_{\mathrm{CM}}\sqrt{s_{0}}}\bigg|_{Q^{2}=0}\displaybreak[0]\nonumber\\
=&\sigma_{T}dn^{\gamma}\bigg|_{Q^{2}=0},\displaybreak[0]
\end{align}
where the contribution of $\sigma_{L}$ and the terms proportional to $Q^{2}$ are neglected in the limit $Q^{2}\rightarrow 0$.
And the general form of the photon spectrum $f_{\gamma}(y)$, which is associated with various particles, reads
\begin{align}\label{fgamma.Gen.}
&f^{\gamma}(y)=\frac{dn^{\gamma}}{dy}=y\int\frac{dQ^{2}}{Q^{2}}\frac{e_{\alpha}^{2}\alpha_{\mathrm{em}}}{2\pi}\rho^{++}\displaybreak[0]\nonumber\\
=&\frac{e_{\alpha}^{2}\alpha_{\mathrm{em}}}{2\pi}\int\frac{dQ^{2}}{Q^{2}}\left\{yC\left(Q^{2}\right)+\left[\frac{2(1-y)}{y}
-\frac{2ym_{\alpha}^{2}}{Q^{2}}\right]D\left(Q^{2}\right)\right\}.\displaybreak[0]\nonumber\\
\end{align}

In the case of coherent-photon emission, Ref.~\cite{Ma:2021lgv} presented a modified photon flux function of proton from Eq.~(\ref{fgamma.Gen.}).
Neglecting the magnetic form factor and adopting the dipole form of electric form factor of proton: $C(Q^{2})=D(Q^{2})=G_{\mathrm{E}}^{2}(Q^{2})$,
and employing the coherent condition $Q^{2}\leq 1/R^{2}_{A}$ ($Q^{2}_{\mathrm{max}}=0.027, y_{\mathrm{max}}=0.16$), one obtains with $a=2m_{p}^{2}/Q^{2}_{\mathrm{max}}$ and $b=2m_{p}^{2}/0.71=2.48$,
\begin{eqnarray}\label{fgamma.MD.}
f_{\mathrm{MD}}^{\gamma}(y)\!\!\!\!&=&\!\!\!\!\frac{\alpha_{\mathrm{em}}}{2\pi}y\left[a-2x+\left(2x+c_{1}\right)d_{1}+\left(2x+c_{2}\right)d_{2}\right.\nonumber\\
&&\!\!\!\!+\left.\left(3x+c_{3}\right)d_{3}+\left(2x+c_{4}\right)d_{4}\right],
\end{eqnarray}
where $x$ depends on $y$,
\begin{eqnarray}\label{x}
x=-\frac{1}{y}+\frac{1}{y^{2}}.
\end{eqnarray}.

Actually, the origin of various widely employed photon spectra is another plane wave form, which is given in Ref.~\cite{Budnev:1974de} and can be written as
\begin{align}\label{fgamma.Gen.V}
dn^{\gamma}(y)&=\frac{e_{\alpha}^{2}\alpha_{\mathrm{em}}}{\pi}\frac{dy}{y}\frac{dQ^{2}}{Q^{2}}\left[\frac{y^{2}}{2}D\left(Q^{2}\right)\right.\displaybreak[0]\nonumber\\
&\left.+\left(1-y\right)\frac{Q^{2}-Q^{2}_{\mathrm{min}}}{Q^{2}}C\left(Q^{2}\right)\right],\displaybreak[0]
\end{align}
this form is achieved from the complete form Eq.~(\ref{fgamma.Gen.}) by assuming that, $Q^{2}_{\mathrm{min}}=y^{2}m_{\alpha}^{2}/(1-y)$, which is the LO term of the following complete expression
\begin{align}\label{Q2lim.}
Q^{2}_{\mathrm{min}}&=-2m_{\alpha}^{2}+\frac{1}{2s_{\alpha b}}\left[\left(s_{\alpha b}+m_{\alpha}^{2}\right)\left(s_{\alpha b}-\hat{s}+m_{\alpha}^{2}\right)\right.\displaybreak[0]\nonumber\\
&\left.-(s_{\alpha b}-m_{\alpha}^{2})\sqrt{(s_{\alpha b}-\hat{s}+m_{\alpha}^{2})^{2}-4s_{\alpha b}m_{\alpha}^{2}}\right].\displaybreak[0]
\end{align}
This approximation is only available when $m_{\alpha}^{2}\ll1~\mathrm{GeV}^2$, however $m_{p}^{2}$ and $m_{Pb}^{2}$ do not satisfy the condition; this is a error source in various spectra.
Especially for lead ion, this approximation cause erroneous results.

Drees and Zeppenfeld (DZ) provided another widely used photon distribution function of proton~\cite{Drees:1988pp, Zhu:2015qoz,Yu:2015kva,Yu:2017rfi,Yu:2017pot,Fu:2011zzf,Fu:2011zzm}, which  is the approximate analytic form of Eq.~(\ref{fgamma.Gen.V}).
Assuming: $Q^{2}_{\mathrm{max}}\rightarrow\infty$, $C\left(Q^{2}\right)=D\left(Q^{2}\right)=G_{\mathrm{E}}^{2}(Q^{2})$, and $Q^{2}-Q^{2}_{\mathrm{min}}\approx Q^{2}$, they obtained
\begin{align}\label{fgamma.DZ.}
f_{\mathrm{DZ}}^{\gamma}(y)&=\frac{\alpha_{\mathrm{em}}}{2\pi}\frac{1+\left(1-y\right)^{2}}{y}\displaybreak[0]\nonumber\\
&\times\left(\ln A-\frac{11}{6}+\frac{3}{A}-\frac{3}{2A^{2}}+\frac{1}{3A^{3}}\right),\displaybreak[0]
\end{align}
where $A=\left(1+0.71~\mathrm{GeV}^{2}/Q^{2}_{\mathrm{min}}\right)$.
$f^{\gamma}_{\mathrm{DZ}}$ properly include electric form factor of proton to describe the situation of the proton as photon emitter.
Because WWA is usually adopted in electroproduction processes, if one directly obtains the spectrum of proton from that of electron by just replacing the $m_{e}$ with $m_{p}$, it would extensively overestimate the cross section.
In Ref.~\cite{Drees:1989vq}, Drees, Ellis, and Zeppenfeld (DEZ) also performed a spectrum of lead ion.
Assuming $y\ll1$, $Q^{2}_{\mathrm{max}}\sim\infty$, $C_{Pb}\left(Q^{2}\right)=0$, and $D_{Pb}\left(Q^{2}\right)\approx\exp\left(-\frac{Q^{2}}{Q^{2}_{0}}\right)$, they achieved
\begin{align}\label{DEZ.fgamma.coh.}
f_{\mathrm{DEZ}}^{\gamma}(y)&=\frac{\alpha_{\mathrm{em}}}{\pi}\left[-\frac{\exp(-Q^{2}_{\mathrm{min}}/Q^{2}_{0})}{y}\right.\displaybreak[0]\nonumber\\
&+\left.\left(\frac{1}{y}+\frac{M^{2}}{Q^{2}_{0}}y\right)\Gamma\left(0,\frac{Q^{2}_{\mathrm{min}}}{Q^{2}_{0}}\right)\right],\displaybreak[0]
\end{align}
where $Q^{2}_{\mathrm{min}}=m^{2}_{Pb}y^{2}$, $\Gamma(a,Q^{2}_{\mathrm{min}}/Q^{2}_{0})=\int_{y}^{\infty}t^{a-1}e^{-t}dt$ is the incomplete Gamma Function.
It should be noticed that, $y\ll1$ means $Q^{2}_{\mathrm{max}}\sim0$, which contradicts with the assumption $Q^{2}_{\mathrm{max}}\sim\infty$.

Based on Eq.~(\ref{fgamma.DZ.}), Nystrand derived a modified photon spectrum of proton which includes the $Q^{2}_{\mathrm{min}}$ term in Eq.~(\ref{fgamma.Gen.V}) and can be presented as~\cite{Nystrand:2004vn}
\begin{align}\label{fgamma.Ny.}
f_{\mathrm{Ny}}^{\gamma}(y)&=\frac{\alpha_{\mathrm{em}}}{2\pi}\frac{1+(1-y)^{2}}{y}\displaybreak[0]\nonumber\\
&\times\left[\frac{A+3}{A-1}\ln A-\frac{17}{6}-\frac{4}{3A}+\frac{1}{6A^{2}}\right].\displaybreak[0]
\end{align}
In addition, the effect of including the magnetic form factor of the proton has been estimated by Kniehl, the final expression $f^{\mathrm{Kn}}(y)$ (Eq.~(3.11) of Ref.~\cite{Kniehl:1990iv}) is too long to include here, but will be discussed further below.

Another most important approach for coherent photon spectrum is the semiclassical impact parameter description, which excludes the hadronic interaction easily.
The calculation of the semiclassical photon spectrum for the case of E1 (electric dipole) excitation is explained in Ref~\cite{CED}, and the result is
\begin{align}\label{fgamma.SC}
f_{\mathrm{SC}}^{\gamma}(y)&=\frac{2Z^{2}\alpha_{\mathrm{em}}}{\pi}\left(\frac{c}{\upsilon}\right)^{2}\frac{1}{y}\bigg\{\xi K_{0}(\xi)K_{1}(\xi)\displaybreak[0]\nonumber\\
&+\frac{\xi^{2}}{2}\left(\frac{\upsilon}{c}\right)^{2}\left[K^{2}_{0}(\xi)-K^{2}_{1}(\xi)\right]\bigg\},\displaybreak[0]
\end{align}
where $\upsilon$ is the velocity of the point charge $Ze$, $K_{0}(x)$ and $K_{1}(x)$ are the modified Bessel functions, and $\xi=\omega b_{\mathrm{min}}/\gamma_{L}v=b_{\mathrm{min}}m_{A}y/v$.
Although the semi-classical photon spectrum is $Q^{2}$-independent, the boundary of $y$ should be constricted by coherence condition~\cite{Baur:2001jj}.

In the case of ultra-incoherent photon emission, Ref.~\cite{Drees:1994zx} presented a photon spectrum inside a quark from Eq.~(\ref{fgamma.Gen.V}). Neglecting the weighting factor in Eqs.~(\ref{CDincoh.pp}) and (\ref{CDUIC}), and setting $Q_{\mathrm{min}}^{2}=1~\mathrm{GeV}^{2}$ and $Q^{2}_{\mathrm{max}}=\hat{s}/4$, one obtains
\begin{align}\label{fgamma.incohI.}
f_{q}^{\gamma}(y)=e_{\alpha}^{2}\frac{\alpha_{\mathrm{em}}}{2\pi}\frac{1+\left(1-y\right)^{2}}{y}\ln\frac{Q_{\mathrm{max}}^{2}}{Q_{\mathrm{min}}^{2}}.\displaybreak[0]
\end{align}
Finally, Brodsky, Kinoshita and Terazawa calculated another important incoherent photon spectrum in Ref.~\cite{Brodsky:1971ud}, which is originally derived for $e$-$p$ scattering,
\begin{align}\label{fgamma.incohBKT}
&f_{\mathrm{BKT}}^{\gamma}(y)\displaybreak[0]\nonumber\\
&=\frac{e_{\alpha}^{2}\alpha_{\mathrm{em}}}{\pi}\bigg\{\frac{1+\left(1-y\right)^{2}}{y}\left(\ln\frac{E}{m}-\frac{1}{2}\right)\displaybreak[0]\nonumber\\
&+\frac{y}{2}\left[\ln\left(\frac{2}{y}-2\right)+1\right]+\frac{\left(2-y\right)^{2}}{2y}\ln\left(\frac{2-2y}{2-y}\right)\bigg\}.\displaybreak[0]
\end{align}

In high-energy physics, the WWA is often used in studying hadronic interactions and heavy-ion collisions:
Wangmei Zha et al. generalized WWA to calculating the cross section and $p_{T}$ distribution of the Breit-Wheeler process in relativistic heavy-ion collisions and their dependence on collision impact parameter $(b)$~\cite{Zha:2018tlq}.
Based on the WWA, theses author also analyzed gluon shadowing in heavy nuclei through Bayesian Reweighting of coherent $J/\psi$ photoproduction in UPCs~\cite{Lu:2025tzs};
ATLAS collaboration studied the exclusive muon pair production in the two-photon process~\cite{ATLAS:2017sfe};
%Drees and Zeppenfeld~\cite{Drees:1988pp} applied the WWA to the production of supersymmetric particles in elastic $e$-$p$ collisions;
d'Enterria and Silveira~\cite{dEnterria:2013zqi} proposed using equivalent photons from colliding lead ions to observe light-by-light scattering experimentally,
later, the direct observations were obtained by the ATLAS~\cite{ATLAS:2019azn} and CMS~\cite{CMS:2018erd} collaborations using lead-lead collisions;
Monte-Carlo event generators, such as Pythia~\cite{Bierlich:2022pfr} and STARlight~\cite{Klein:2003vd}, also utilize the WWA.
Generally, particle production in hadronic collisions is often modelled using WWA~\cite{Klein:2016yzr,Klein:2003vd}.
Although great development has been achieved, the properties of WWA in inelastic heavy quarkonium photoproduction in heavy-ion collisions are often neglected, and the imprecise statements were given~\cite{Zhu:2015via, Zhu:2015qoz, Fu:2011zzm, Fu:2011zzf, Chin.Phys.C_36_721, Yu:2015kva, Yu:2017rfi, Yu:2017pot, Drees:1989vq, Drees:1988pp, Frixione:1993yw, Nystrand:2004vn, Kniehl:2001tk, Kniehl:1990iv, Yang:2019lls, Wu:2020ujf}.
Especially in the ultrarelativistic heavy-ion collisions at LHC energies, WWA becomes deterministic to the accuracy of describing photoproductions, since $f_{\gamma}\propto Z^{2}\ln\sqrt{s}/m$, $\sqrt{s}$ and $Z^{2}$ are the very large factors.
Therefore, we will comprehensively analyze the WWA in the mentioned case, and estimate the inaccuracies in above spectra.

\section{Numerical results}
\label{Numerical results}

\begin{figure*}[htbp]
\setlength{\abovecaptionskip}{1mm}
  \centering
  \includegraphics[width=0.92\textwidth]{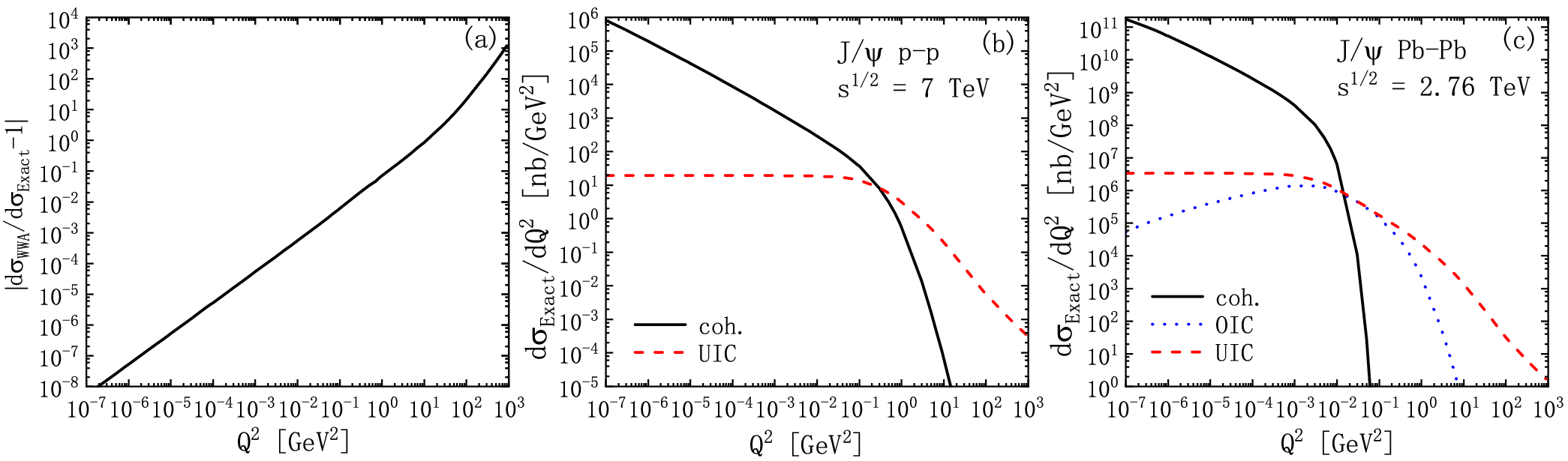}
  \caption{$Q^{2}$ distribution of the $J/\psi$ photoproduction at LHC energies.
  (a): relative error of the WWA result with respect to the exact one.
  (b), (c): exact results of the $Q^{2}$-dependent differential cross sections in $p$-$p$ and $Pb$-$Pb$ collisions, respectively.
  Black solid line---coherent-photon emission [coh.(dir.+res.)].
  Red dashed line---ultra-incoherent photon emission [UIC (dir.+res.)].
  Blue dotted line---ordinary-incoherent photon emission [OIC (dir.+res.)].
  }
  \label{fig:Q2}
\end{figure*}

\begin{figure*}[htbp]
\setlength{\abovecaptionskip}{1mm}
  \centering
  \includegraphics[width=0.92\textwidth]{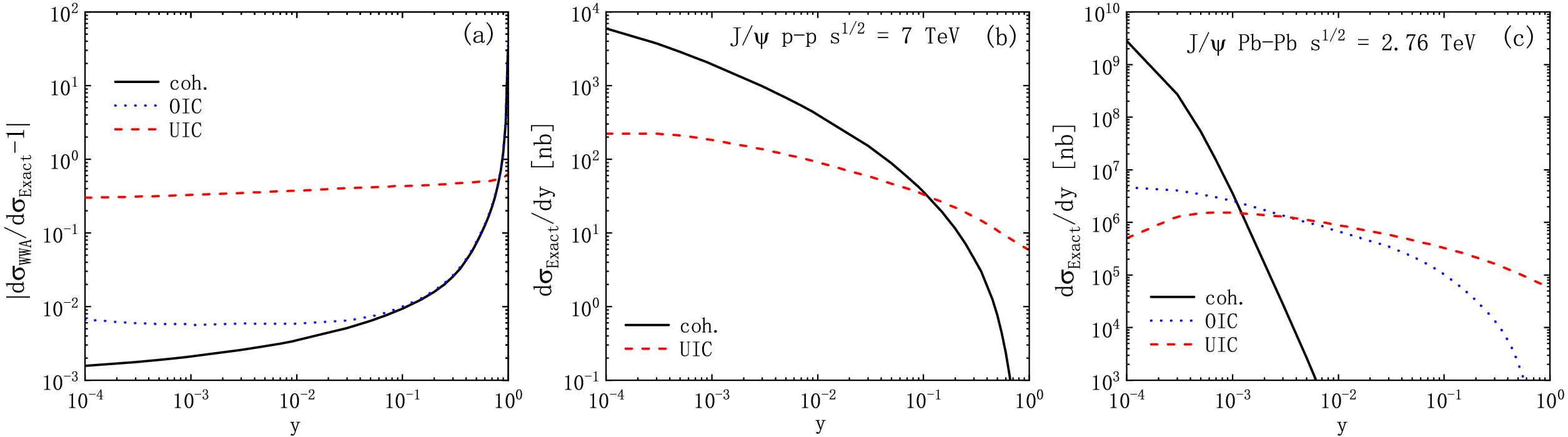}
  \caption{Same as Fig.~\ref{fig:Q2} but for $y$ distribution.
  %Black solid line---coherent-photon emission [coh.(dir.+res.)].
  %Red dashed line---ultra-incoherent photon emission [UIC (dir.+res.)].
  %Blue dotted line---ordinary-incoherent photon emission [OIC (dir.+res.)].
  }
  \label{fig:y}
\end{figure*}

\begin{figure*}[htbp]
\setlength{\abovecaptionskip}{1mm}
  \centering
  \includegraphics[width=0.92\textwidth]{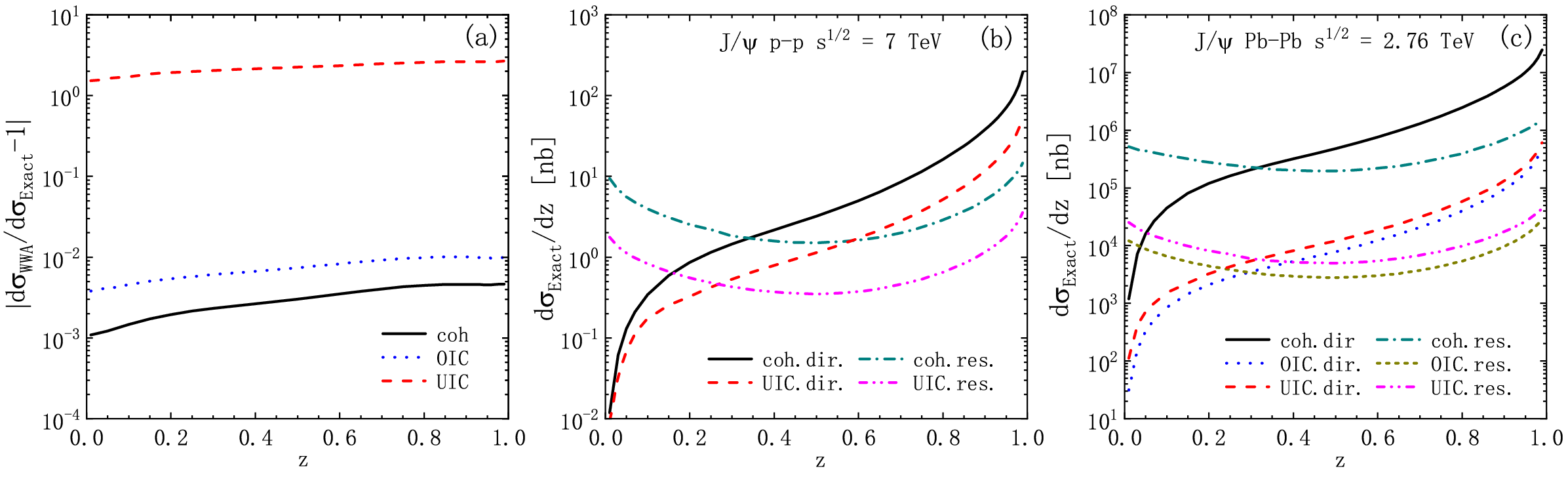}
  \caption{(a): relative error of the WWA result with respect to the exact one.
  (b), (c): exact results of $d\sigma/dz$ in $p$-$p$ and $Pb$-$Pb$ collisions.
  Black solid and dark cyan dot-dashed lines denote the coherent-direct and resolved photon emissions, respectively.
  Red dashed and magenta dot-dot dashed lines denote the ultra-incoherent direct and resolved photon emissions, respectively.
  Blue dotted and dark yellow short dashed lines denote the ordinary-incoherent direct and resolved photon emissions, respectively.
  }
  \label{fig:z}
\end{figure*}

Now, we are ready to show the numerical results.
In the calculations, we adopt an approximation $m_{q}=M_{H}/2$ for the quark mass, where $M_{H}$ is the mass of heavy quarkonium.
All the masses are taken from PDG~\cite{Agashe:2014kda}.
We use $m_{p}=0.938~\mathrm{GeV}$, $\alpha_{\mathrm{em}}=1/137.036$, $p_{T\mathrm{min}}=M_{H}$ and the two-loop QCD coupling constants $\alpha_{s}$ with $n_{f}=3$ and $\Lambda=0.2~\mathrm{GeV}$~\cite{Ma:2015ykd}.
We adopt MSHT20 LO (NLO) set for PDFs with $n_{f}=3$~\cite{Bailey:2020ooq}, and the factorization scale $\mu=\sqrt{M_{H}^{2}+Q^{2}}$.
The LDMEs and the full kinematical relations are summarized in Appendices.

\begin{figure*}[htbp]
\setlength{\abovecaptionskip}{1mm}
  \centering
  \includegraphics[width=0.92\textwidth]{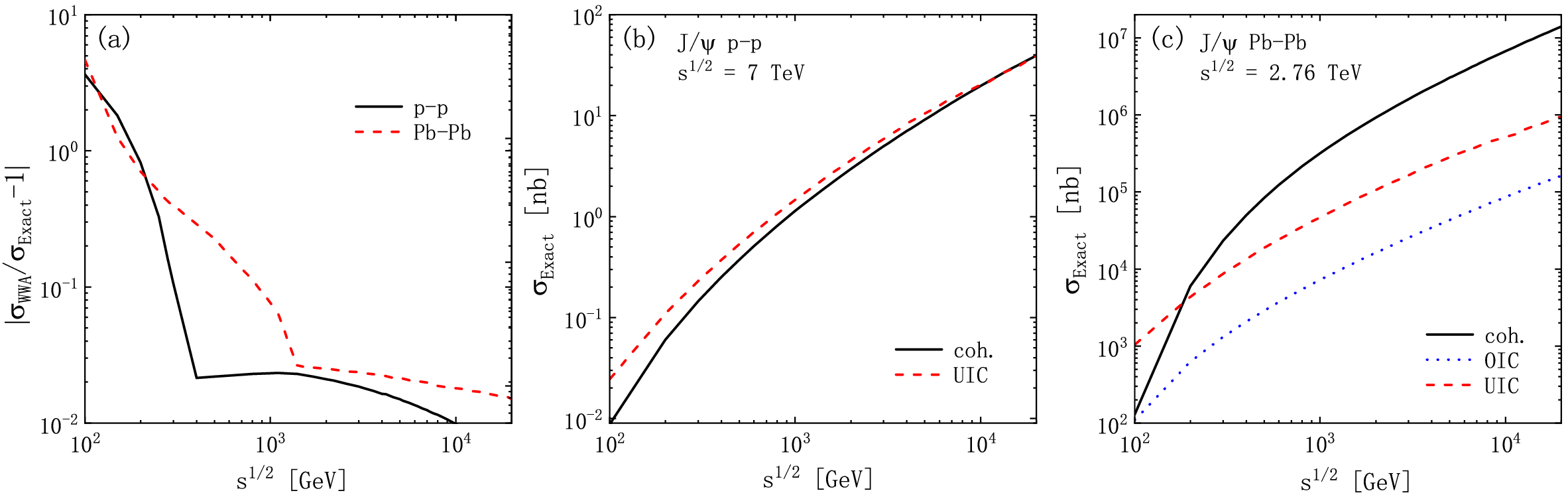}
  \caption{Same as Fig.~\ref{fig:Q2} but for the total cross sections as a function of $\sqrt{s}$.
  }
  \label{fig:s}
\end{figure*}

First of all, we choose the $J/\psi$ as an example, to comprehensively study the properties of WWA in inelastic heavy quarkonium photoproduction in heavy-ion collisions.
Therefore, the distributions in $Q^2$, $y$, $z$, and $\sqrt{s}$ are plotted in Figs.~\ref{fig:Q2}-\ref{fig:s}, where all the results are the sum of direct and resolved contributions, and do not include the fragmentation contributions.
The left panels of Figs.~\ref{fig:Q2}-\ref{fig:s} show the relative errors with respect to the exact results, the central and right panels show the exact results of cross sections in $p$-$p$ and $Pb$-$Pb$ collisions, respectively.

In Fig.~\ref{fig:Q2}, there is only one curve in panel $(a)$, since the curves of coh., OIC and UIC are consistent with each other.
We observe that the relative error can be neglected in small $Q^{2}$ region, but becomes non-negligible at $Q^{2}\gtrsim1~\mathrm{GeV}^{2}$ and shows the rapid growing at high $Q^{2}$ region.
At $Q^{2}=1~\mathrm{GeV}^{2}$, the relative error is about $7\%$; at $Q^{2}=10~\mathrm{GeV}^{2}$, the error reaches up to $86\%$.
Therefore, WWA has a good accuracy only in very small $Q^{2}$ domain where exact treatment can nicely recover to WWA.
But in the large $Q^{2}$ region, WWA becomes inapplicable where the deviation is prominent, especially when $Q^{2}>1~\mathrm{GeV}^{2}$.

In panels $(b)$ and $(c)$, the coherent and ultra-incoherent photon emissions dominate the small and large $Q^{2}$ regions, respectively.
They become comparable at $Q^{2}=10^{-1}$ and $10^{-2}~\mathrm{GeV}^{2}$ in $p$-$p$ and $Pb$-$Pb$ collisions, respectively.
We notice that, except for the region $10^{-2}<Q^{2}<10^{-1}~\mathrm{GeV}^{2}$, the contribution of ordinary-incoherent photon emission can be neglected safely comparing with other two channels.
According to the views derived from panel $(a)$, one can deduce that WWA is valuable in coherent and ordinary incoherent-photon emissions.
Especially in $Pb$-$Pb$ collisions, the WWA has a much higher accuracy in coherent process which quickly approaches to zero before $Q^{2}=10^{-2}~\mathrm{GeV}^{2}$, this effectively avoid WWA error.
However, WWA is not valid for ultra-incoherent photon emission, which concentrates on the large $Q^{2}$ domain where the WWA error is prominent.

In Fig.~\ref{fig:y}, the results are expressed as a function of $y$.
In panel (a), the curves of coh. and OIC share the same trend, and are consistent with each other in the large $y$ domain.
We find that $y=0.1$ is a key point: when $y<0.1$ the WWA results nicely agree with the exact ones for both channels, and WWA has a better accuracy in coherent process rather than OIC one (at $y=10^{-4}$, the deviations are about $1.6\text{\textperthousand}$ and $6.7\text{\textperthousand}$ in each case, respectively);
but when $y>0.1$, the relative errors become prominent, especially the curves show a pronounced peak near $y=1$.
Therefore, WWA is only effective when $y<0.1$ and it has evident error at large values of $y$.
Additionally, the relative error of UIC is prominent (about $0.3$-$0.6$) in the whole $y$ regions.

\begin{table*}[htbp]\footnotesize
\renewcommand\arraystretch{1.1}
\centering
\caption{\label{Total.CS.coh.pp} $J/\psi$ photoproduction in the channel of coherent-photon emission in $p$-$p$ collisions [$7~\mathrm{TeV}$].}
\begin{tabular}{L{2.0cm}R{2.0cm}R{2.0cm}R{2.0cm}R{2.0cm}R{2.0cm}R{2.0cm}}
\hline\noalign{\smallskip}
  coh.(dir.+res.)   &  Exact  & $f_{\mathrm{DZ}}^{\gamma}$ & $f_{\mathrm{Ny}}^{\gamma}$  &  $f_{\mathrm{Kn}}^{\gamma}$  &  $f_{\mathrm{SC}}^{\gamma}$  &  $f_{\mathrm{MD}}^{\gamma}$ \\
\noalign{\smallskip}\hline\noalign{\smallskip}
  $\sigma~~[\mathrm{nb}]$    & 22.4241 & 43.3561 & 35.1059 & 39.9058 & 20.3327 & 22.6512 \\
  $\delta\tnote{a}~~[$\%$]$  & 0.00    & 93.35   & 56.55   & 77.96   & 9.33    & 1.01    \\
\hline\noalign{\smallskip}
\end{tabular}
\begin{tablenotes}
\scriptsize
\item[a] Relative error with respect to the exact result: $\delta=|\sigma/\sigma_{\mathrm{Exact}}-1|$.
\end{tablenotes}
\end{table*}

In panels (b) and (c), the coherent processes dominate the very small $y$ regions, this behaviour guarantees the accuracy of WWA.
Conversely, the UIC contributions are predominantly from the regions $y>10^{-1}$ and $y>10^{-3}$ in $p$-$p$ and $Pb$-$Pb$ collisions, respectively.
Comparing with the views derived from panel (a), we can verify that the validity range of WWA is compatible with coh. and OIC processes.
Particularly in $Pb$-$Pb$ collisions, the WWA has a better accuracy in coherent process, since its contribution fall off far more rapidly.
However, in UIC process, WWA is inapplicable in the entire $y$ regions.

In Fig.~\ref{fig:z}, we provide the cross section $d\sigma/dz$ vs $z$.
The panel (a) explicitly shows us that, WWA is applicable in coh. and OIC processes in the entire $z$ range, and has the highest accuracy in coh. process (the deviation is about $1.1\text{\textperthousand}-4.7\text{\textperthousand}$).
Whereas the deviation of UIC process reaches up to $153\%$-$268\%$ in the whole $z$ regions.
In panels (b) and (c), we find that the $z$ behaviour of each channel is quite different.
Firstly, the resolved contributions dominate the lower $z$ region and become smaller than those of direct contributions when $z>0.3$; this feature agrees with the traditional perspective that the resolved process contributes appreciably only at  $z\leq0.3$~\cite{Kniehl:1998qy}.
Secondly, the results are divergent near the endpoint $z=1$, which are mainly from the color-octet $^{1}S_{0}$ and $^{3}P_{J}$ processes.
The reason is that the NRQCD prediction breaks down and the color-octet channels exhibit collinear singularities in the region of $z\lesssim1$, where diffractive production takes place.
In order to screen the collinear singularities and suppress the elastic production, the traditional way is to impose the cuts: $z<0.9$, $M_{X'}>10~\mathrm{GeV}$ and $p_{T}>1~\mathrm{GeV}$ or $M_{J/\psi}$ (actually, if $p_{T~\mathrm{min}}\neq0$, $z_{\mathrm{max}}$ will naturally less than one).
Another possibility to suppress the elastic production at $z\lesssim1$ would be to require that $Q^{2}$ be sufficiently large~\cite{Kniehl:2001tk}.
However, then also the bulk of the inelastic contribution would be sacrificed.
Finally, the coherent contributions are larger than those of UIC processes in the whole $z$ regions.
Especially in $Pb$-$Pb$ collisions, coherent contribution dominates the entire $z$ regions about an order of magnitude than other two channels.
Since the equivalent photon flux scales as $Z^{2}$, which is a large enhancement factor for the cross section.

In Fig.~\ref{fig:s}, we show the total cross sections distributed in $\sqrt{s}$.
In panel (a), the curves of coh. and OIC are consistent with each other.
We do not show the curve of UIC process, where the deviation is $0.38$-$0.49$ in the whole $\sqrt{s}$ ranges.
It can be seen that the curves show a pronounced rising when $\sqrt{s}<400$ and $1400~\mathrm{GeV}$ in $p$-$p$ and $Pb$-$Pb$ collisions, respectively.
And they slowly decreased with increasing $\sqrt{s}$.
The trend of curves is similar to the results of Kniehl~\cite{Kniehl:1990iv}.
Therefore, WWA has the significant errors in small $\sqrt{s}$ domain (RHIC energies), and has a good accuracy at high energies (LHC energies).
In panels (b) and (c), the coherent process is slightly smaller than UIC one in $p$-$p$ collisions;
while the situation is opposite in $Pb$-$Pb$ collisions, where the coherent process starts to play a fundamental role (it is about one and two orders of magnitude  larger than OIC and UIC processes, respectively).
The reason is that the coherent process scale with $Z^{2}$, whereas the OIC and UIC processes scale approximately with $Z$ and $N_{A}$, respectively.
As for $Z\gg1$, the coherent part dominates the production processes.
Based on the views derived from the left panel and previous discussions, we can deduce the further conclusion that WWA can reach the high accuracy in high energy range and in $Pb$-$Pb$ collisions, where the UIC contribution is greatly suppressed;
however it is not a good approximation in $p$-$p$ collisions, where the UIC contribution is comparable with the coherent one.

In the comprehensive discussions of the results displayed in above four figures [Figs.~\ref{fig:Q2}-\ref{fig:s}], we obtained the validity scopes of WWA.
Therefore, in deriving of the equivalent photon spectra based on WWA, the kinematical boundaries are crucial to the accuracy.
Now, we display the total cross sections in Tables~\ref{Total.CS.coh.pp}-\ref{Total.CS.UIC}, to discuss the accuracies and its sources of the widely employed photon spectra mentioned in Section~\ref{WWA}.
In $p$-$p$ collisions [Table~\ref{Total.CS.coh.pp}], the relative error of $f^{\gamma}_{\mathrm{DZ}}$ is the largest,
since the integration is performed in the entire kinematical allowed regions: $Q^{2}_{\mathrm{max}}=\infty$ and $y_{\mathrm{max}}=1$, we know that from Figs.~\ref{fig:Q2} and \ref{fig:y}, the $Q^{2}>1~\mathrm{GeV}^{2}$ and $y>0.1$ domains are able to give a large fictitious contribution.
For the spectrum $f^{\gamma}_{\mathrm{Ny}}$, the deviation has a obvious reduction compared to $f^{\gamma}_{\mathrm{DZ}}$,
since it includes the $Q^{2}_{\mathrm{min}}$ term in Eq.~(\ref{fgamma.Gen.V}) which is omitted in $f^{\gamma}_{\mathrm{DZ}}$.
Actually this term is inversely proportional to $Q^{4}$ and thus has the noticeable effect in small $Q^{2}$ region, which can not be neglected when performing the photon spectra;
the perspective agrees with Kniehl~\cite{Kniehl:1990iv}.
The relative error of $f^{\gamma}_{\mathrm{Kn}}$ is higher than that of $f^{\gamma}_{\mathrm{Ny}}$ but lower than that of $f^{\gamma}_{\mathrm{DZ}}$, since the magnetic form factor of proton is also included in this form,
which effects only the large $Q^{2}$ range and should be essentially excluded~\cite{Ma:2019mwr}.
Finally, the modified proton spectra $f^{\gamma}_{\mathrm{MD}}$ nicely agrees with the exact ones ($\delta\sim1\%$).
There are two reasons: $f^{\gamma}_{\mathrm{MD}}$ is derived from the complete form Eq.~(\ref{fgamma.Gen.}) which includes the $Q^{2}_{\mathrm{min}}$ term and properly excludes the effects of magnetic form factor; it adopts the coherence condition, which means that the wavelength of the photon is larger than the size of the nucleus, and the charged constituents inside the nucleus should act coherently.
This condition effectively cut the WWA errors.
There are also other limitations which can reach the such high accuracy, the key point is that, in most of the physically interesting cases a dynamical cut off exists, such that, the photo-absorption cross sections differ slightly from their values on the mass shell. The details can be found in Ref.~\cite{Budnev:1974de}.

\begin{table}[htbp]\footnotesize
\renewcommand\arraystretch{1.1}
\centering
\caption{\label{Total.CS.coh.PbPb}Same as Table~\ref{Total.CS.coh.pp} but in $Pb$-$Pb$ collisions [$2.76~\mathrm{TeV}$].}
\begin{tabular}{L{2cm}R{1.7cm}R{1.7cm}R{1.7cm}}
\hline\noalign{\smallskip}
  coh.(dir.+res.)  &  Exact  & $f_{\mathrm{DEZ}}^{\gamma}$ & $f_{\mathrm{SC}}^{\gamma}$ \\
\noalign{\smallskip}\hline\noalign{\smallskip}
  $\sigma~[\mathrm{mb}]$   & 2.4854  & 18.9783 & 1.9274 \\
  $\delta~[$\%$]$          & 0.00    & 663.59  & 22.45  \\
\noalign{\smallskip}\hline
\end{tabular}
\end{table}

In $Pb$-$Pb$ collisions [Table~\ref{Total.CS.coh.PbPb}],
the deviation of $f^{\gamma}_{\mathrm{DEZ}}$ reaches up to $680.19\%$, since $f^{\gamma}_{\mathrm{DEZ}}$ is constructed on a contradictory assumption:
$Q^{2}_{\mathrm{max}}\sim\infty$ and $y\ll1$ (actually, $Q^{2}_{\mathrm{max}}\sim\infty$ is equivalent to $y_{\mathrm{max}}\sim1$).
For $f^{\gamma}_{\mathrm{SC}}$, it roughly agrees with the exact ones in both $p$-$p$ and $Pb$-$Pb$ collisions, but the deviations still can not be neglected.
Since $f^{\gamma}_{\mathrm{SC}}$ is calculated from the semiclassical impact parameter description, where the coherence condition is used.
One should be careful that when using $f^{\gamma}_{\mathrm{SC}}$, setting $y_{\mathrm{max}}=1$ will cause the erroneous results.

\begin{table}[htbp]\footnotesize
\renewcommand\arraystretch{1.2}
\centering
\caption{\label{Total.CS.UIC}Same as Table~\ref{Total.CS.coh.pp} but for ultra-incoherent photon emission in $p$-$p$ [$7~\mathrm{TeV}$] and $Pb$-$Pb$ [$2.76~\mathrm{TeV}$] collisions.}
\begin{tabular}{L{1.8cm}R{1.75cm}R{1.75cm}R{1.75cm}}
\hline\noalign{\smallskip}
  $\sigma_{\mathrm{UIC (dir.+res.)}.}$  &  Exact  & $f_{q}^{\gamma}$ & $f_{\mathrm{BKT}}^{\gamma}$ \\
\noalign{\smallskip}\hline\noalign{\smallskip}
  $\sigma_{pp}\ [\mathrm{nb}]$    & 24.2773 & 39.2290 & 101.3291 \\
  $\delta_{pp}\ [$\%$]$           & 0.00    & 61.59   & 317.38   \\
  $\sigma_{PbPb}\ [\mathrm{mb}]$  & 0.2494  & 0.6023  & 0.7254   \\
  $\delta_{PbPb}\ [$\%$]$         & 0.00    & 141.51  & 190.87   \\
\noalign{\smallskip}\hline
\end{tabular}
\end{table}

In the case of ultra-incoherent photon emission [Table~\ref{Total.CS.UIC}], the derivations are prominent and turn to much serious in $Pb$-$Pb$ collisions.
This quantitatively verifies the inapplicability of WWA in UIC process.
In particular, the errors of $f^{\gamma}_{\mathrm{BKT}}$ are the largest, since it is originally derived from $e$-$p$ scattering, but is directly expanded to describe the probability of finding a photon in any relativistic fermion~\cite{Kniehl:1990iv,Yu:2015kva,Yu:2017pot}, this will overestimate the cross sections.
Therefore, the UIC process should be treated in exact treatment, and  the results in Refs.~\cite{Zhu:2015via,Zhu:2015qoz,Fu:2011zzm,Fu:2011zzf,Chin.Phys.C_36_721,Yu:2015kva,Yu:2017rfi,Yu:2017pot,Drees:1989vq,Drees:1988pp,Frixione:1993yw,Nystrand:2004vn,Kniehl:2001tk,Kniehl:1990iv,Yang:2019lls,Wu:2020ujf}
are not accurate enough, where the mentioned spectra are adopted and the serious double counting exists.

\begin{table}[htbp]\footnotesize
\renewcommand\arraystretch{1.2}
\centering
\caption{\label{Total.CS.Exact}Exact results of the $J/\psi$ and $\Upsilon(1S)$ photoproductions in $p$-$p$ [$7~\mathrm{TeV}$] and $Pb$-$Pb$ [$2.76~\mathrm{TeV}$] collisions.
  All the results are the sum of the direct and resolved contributions, and do not include fragmentation contributions.}
\begin{tabular}{L{1.2cm}R{1.5cm}R{1.2cm}R{1.2cm}R{1.2cm}R{1.2cm}}
\hline\noalign{\smallskip}
  Exact                              &  Quarkonium    & coh.    & UIC     & Total   & Total (NWF)\\
\noalign{\smallskip}\hline\noalign{\smallskip}
  $\sigma_{pp}\ [\mathrm{nb}]$       & $J/\psi$       & 22.4241 & 24.2773 & 46.7014 & 103.9845  \\
  $\sigma_{PbPb}\ [\mathrm{mb}]$     & $J/\psi$       & 2.4854  & 0.2494  & 2.7348  & 3.1055    \\
  $\sigma_{pp}\ [\mathrm{pb}]$       & $\Upsilon(1S)$ & 13.6794 & 26.9652 & 40.6446 & 82.6189   \\
  $\sigma_{PbPb}\ [\mathrm{\mu b}]$  & $\Upsilon(1S)$ & 0.7345  & 0.2113  & 0.9458  & 1.1262    \\
\noalign{\smallskip}\hline
\end{tabular}
\end{table}

\begin{figure*}[htbp]
\setlength{\abovecaptionskip}{1mm}
  \centering
  \includegraphics[height=10cm,width=13cm]{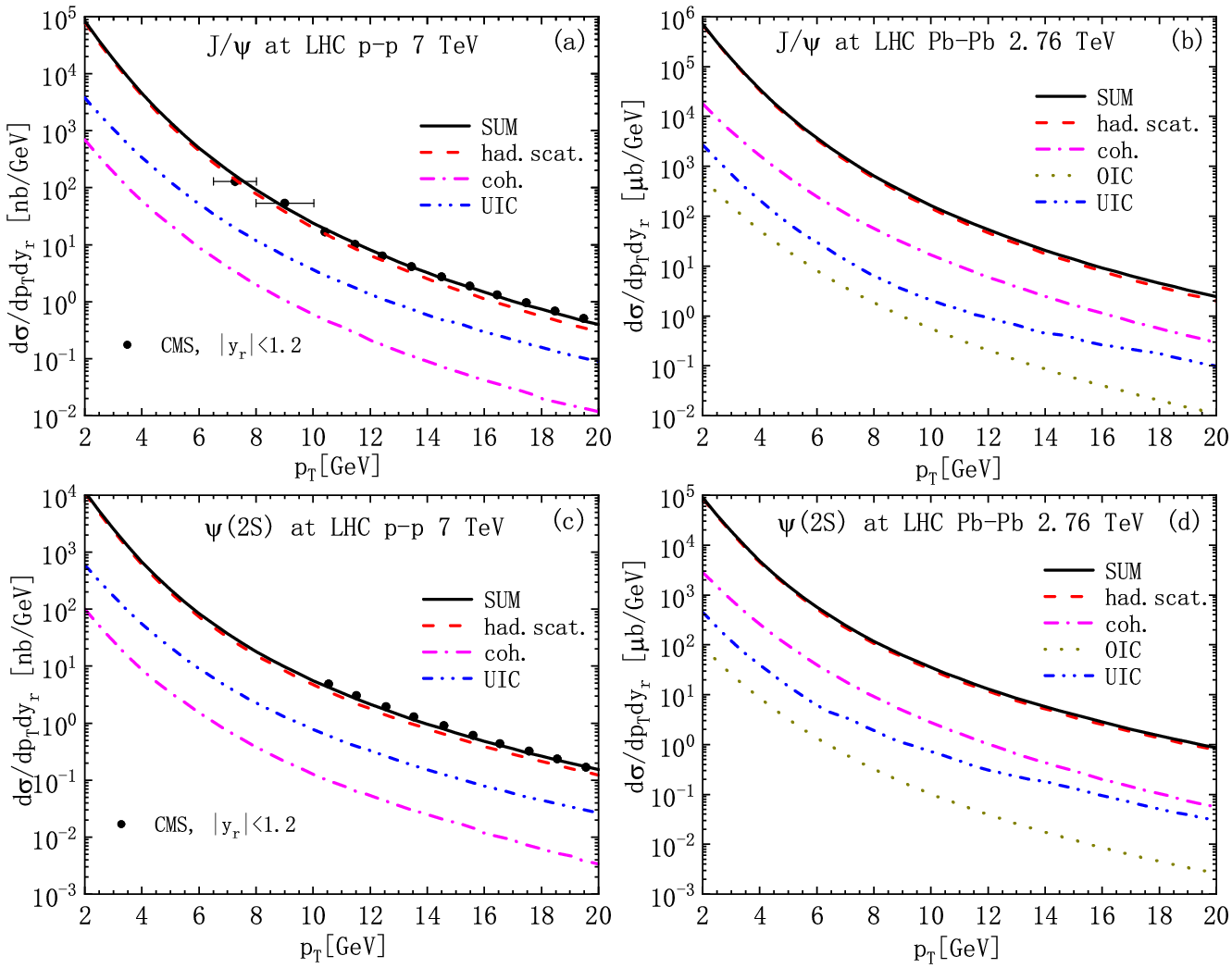}
  \caption{The $p_{T}$ distribution of $J/\psi$ and $\psi(2S)$ photoproductions in $p$-$p$ and $Pb$-$Pb$ collisions.
  Red dashed line denotes the initial partons hard scattering (had.scat.).
  Magenta dot-dashed line is for coherent-photon emissions [coh.(dir.+res.)].
  Blue dot-dot dashed line is for ultra-incoherent photon emissions [UIC (dir.+res.)].
  Dark yellow dotted line is for ordinary-incoherent photon emissions [OIC (dir.+res.)].
  Black solid line is for the sum of had.scat. and photoproduction processes.
  The $J/\psi$ and $\psi(2S)$ data are from the CMS collaboration~\cite{CMS:2015lbl}.
  The rapidity $y_{r}$ is integrated over the experimental range $|y_{r}|<1.2$.}
  \label{fig:dPT.Charmonium}
\end{figure*}

\begin{figure*}[htbp]
\setlength{\abovecaptionskip}{1mm}
  \centering
  \includegraphics[height=10.3cm,width=16.5cm]{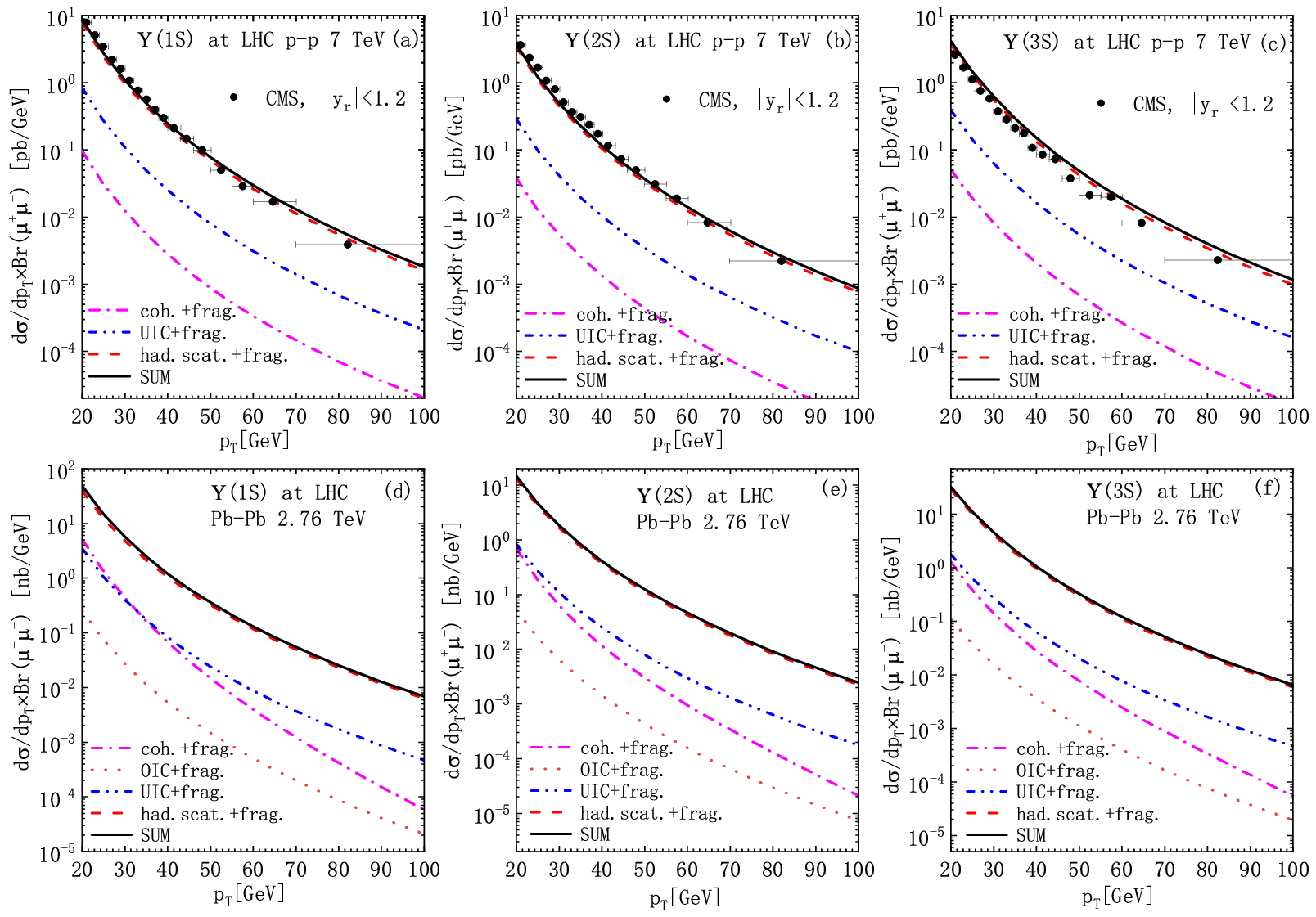}
  \caption{Same as Fig~\ref{fig:dPT.Charmonium}, but for $\Upsilon(nS)$ productions, all the results are the sum of direct and fragmentation $\Upsilon(nS)$ productions.
  The branching fractions are $B_{r}(\Upsilon(1S)\rightarrow\mu^{+}\mu^{-})=2.48\%$, $B_{r}(\Upsilon(2S)\rightarrow\mu^{+}\mu^{-})=1.93\%$,
  and $B_{r}(\Upsilon(3S)\rightarrow\mu^{+}\mu^{-})=2.18\%$.
  The $\Upsilon(nS)$ data are from the CMS collaboration~\cite{CMS:2015xqv}.
  The rapidity $y_{r}$ is integrated over the experimental range $|y_{r}|<1.2$.}
  \label{fig:dPT.Bottonium}
\end{figure*}

Next, in Table~\ref{Total.CS.Exact}, we estimate the contributions of the UIC channel to the heavy quarkonium photoproduction, and discuss the double counting encountered in most of the works when the different photon emissions are considered at the same time.
In $p$-$p$ collisions, UIC result is comparable with coh. one for $J/\psi$, and becomes about two times larger than coh. one for $\Upsilon(1S)$.
In $Pb$-$Pb$ collisions, the ratio of UIC to the coh. one is about $10\%$ for $J/\psi$, and about $29\%$ for $\Upsilon(1S)$.
Therefore, the UIC channel plays the very important role in $p$-$p$ collisions, and can also provide the meaningful contribution in $Pb$-$Pb$ collisions.
In addition, the relative contribution of UIC channel to the inelastic heavy quarkonium photoproduction becomes much more obvious along with the increasing quarkonium mass.

On the other hand, Table~\ref{Total.CS.Exact} also shows us that the double counting is serious.
The total results with no weighting factor (NWF) are much larger than those of exact ones.
These large fictitious contributions are mainly from the lower $Q^{2}$ domain.
The traditional way of excluding these unphysical results is to utilize a cutoff $Q^{2}>1~\mathrm{GeV^{2}}$ when performing the equivalent photon spectra.
Such as the calculation of $f_{q}^{\gamma}$ [Eq.~\ref{fgamma.incohI.}], the cutoff is adopted to replace the remaining probability factor ``$1-w_{\mathrm{coh}}$".
However, we can see that the corresponding errors in Table~\ref{Total.CS.UIC} are still evident.
Therefore, when the different photon emissions are considered simultaneously,  each channel should be multiplied by the probability or weighting factor to avoid double counting.

Now, we turn to study the contribution of photoproduction processes to the heavy quarkonium productions in $p$-$p$ and $Pb$-$Pb$ collisions at LHC energies.
Before the discussion, there is one critical issue must be addressed: heavy quarkonium productions in hadronic collisions (especially $Pb$-$Pb$) are spatially limited, with a collision parameter range within $2R_{A}$ ($R_{A}=A^{1/3}1.2~ \textrm{fm}$ is the size of the nucleus).
When considering the contribution of photonproduction to the hadronic process, the equivalent photon flux should be integrated to $b_{\textrm{max}}=2R_{A}$.
Integrating to infinity artificially enhances the cross section by including unphysical contributions from non-overlapping nuclei at large, violating the spatial constraints of the collision geometry.
However, in present paper, we performed the calculations and developed our formalism based on the parameterized form factor method, which do not include impact parameter $b$ explicitly (where the form factor is derived from the Fourier transform of the charge density $\rho$ of the nucleus and the approximate plane waves are used).
For the consistency of our formalism, we adopt the empirical approximation to include this truncation affects, which is also used in the literature~\cite{Ma:2018zzq, Baur:2001jj, CED}.
Where the effect of $b_{\textrm{max}}=2R_{A}$ is appear in the cut $Q^{2}_{\textrm{min}}$, the details see Appendix~\ref{FKR}.
The strict method to incorporate $b$ is the semiclassical impact parameter description, which is more appropriate to calculate the case of heavy ions collisions and often used in the study of UPCs (recently, Wangmei Zhang et al. developed a spatially-dependent photon flux distribution and established a more precise relationship between the photon transverse momentum distribution and collision impact parameter~\cite{Wu:2024jqn}).

In Figs.~\ref{fig:dPT.Charmonium} and \ref{fig:dPT.Bottonium}, we adopt the exact treatment to plot the $p_{T}$ distributions of charmonium ($J/\psi$, $\psi(2S)$) and bottomonium ($\Upsilon(nS)$) photoproductions, respectively.
Where all the results are the sum of the direct and resolved photon contributions.
For completely showing the photoproduction contribution and studying the $p_{T}$ behaviour of each photon source, we plot the Figs.~\ref{fig:dPT.Charmonium} and \ref{fig:dPT.Bottonium} in different $p_{T}$ ranges which compensate with each other.
In Fig.~\ref{fig:dPT.Charmonium}, we plot the $p_{T}$ distribution of charmonium in the range of $2~\textrm{GeV}<p_{T}<20~\textrm{GeV}$.
Fragmentation mechanism is unreliable at this $p_{T}$ range, because the LP and NLP contributions is valid only for $p_{T}\gg m_{H}$.
(Authors in Ref.~\cite{Bodwin:2015iua,Bodwin:2014gia} have shown that the NLO LP SDC are very different to the fixed-order SDC accurate through NLO at small $p_{T}$.
They suggested that a criterion $p_{T}>3M_{J/\psi}\simeq10~\textrm{GeV}$ to ensure the validity of fragmentation mechanism and suppress NLP contributions.)
In order to consider the NLO effects, the LO results are multiplied by the $K$ factor in Ref.~\cite{Butenschoen:2010rq}, we choose $K=1.8$ and $1.3$ for direct and resolved contributions, respectively.
The spectra of photoproductions are compared to the hard scattering of initial partons.
In panels (a) and (c), we compare the results with data of CMS Collaboration~\cite{CMS:2015lbl}.
At very small $p_{T}$, the color-singlet and color-octet distributions are corrupted by collinear divergences and soft gluon effects~\cite{Cho:1995vh}.
Since the intrinsic motion of incident partons renders the differential cross section uncertain for $p_{T}\leq2~\mathrm{GeV}$, our predictions only concern the range   $p_{T}\geq2~\mathrm{GeV}$.

In $p$-$p$ collisions [panels (a), (c)], the UIC processes are larger than the coh. ones, this verifies again the significance of UIC channel in $p$-$p$ collisions.
However, the coh. contributions are still comparable with UIC ones ($d\sigma_{\mathrm{coh.}}/d\sigma_{\mathrm{UIC}}\sim0.18-0.13 $ in the $p_{T}$ ranges considered).
This is different to the results in Ref.~\cite{Fu:2011zzm}, where the coh. contributions are neglected.
In $Pb$-$Pb$ collisions [panels (b), (d)], the coh. processes are larger than the UIC ones, since they scale with $Z^{2}$, whereas the UIC ones scale with $N_{A}$.
We disagree with the results in Refs.~\cite{Yu:2017pot,Yu:2015kva} where the situation is opposite to us, the UIC contributions are about an order of magnitude larger than coh. ones.
Finally, we notice that the charmonium photoproduction processes give the valuable corrections to the had.scat. when $p_{T}>4~\mathrm{GeV}$, and the corrections become more obvious along with the increasing $p_{T}$.
And compared to $p$-$p$ collisions, the corrections are more obvious in $Pb$-$Pb$ collisions.
This is also different to the the results in Refs.~\cite{Yu:2017pot,Yu:2015kva} where the contribution of photoproduction even larger than the hadronic process.
The reason is that they do not consider the cut $b_{\textrm{max}}=2R_{A}$, artificially enhancing the cross section by including unphysical contributions from non-overlapping nuclei at large.
This cut will exclude the large contribution from small $Q^{2}$ region [see Fig.~\ref{fig:Q2}], and causing the coherent contribution becomes much smaller (for incoherent channel, this cut only cause the small effect).

In Fig.~\ref{fig:dPT.Bottonium}, we plot the exact results of bottomonium ($\Upsilon(nS)$) photoproduction and fragmentation processes.
For avoiding double counting, we adopt the BCCKL method where the matching rule between fixed-order and fragmentation has been performed~\cite{Bodwin:2015iua,Bodwin:2015yma,Bodwin:2014gia}.
Since the fragmentation bottomonium productions are untrustworthy throughout the $p_{T}<15~\mathrm{GeV}$ region~\cite{Cho:1995vh}, and in order to suppress the next to leading power (NLP) contributions.
We present our theoretical predictions for the values of $p_{T}>20~\mathrm{GeV}$.
In upper panels, the exact results are compared to the CMS data~\cite{CMS:2015xqv}.
In the case of $p$-$p$ collisions [panels (a)-(c)], the differences between UIC processes and coh. ones become much larger than those of $J/\psi$ productions in Fig.~\ref{fig:dPT.Charmonium}.
In the case of $Pb$-$Pb$ collisions [panels (d)-(f)], the UIC channel becomes larger than the coh. one at sufficiently large $p_{T}$ ranges.
This feature is opposite to the situation in panels (b), (d) of Fig.~\ref{fig:dPT.Charmonium}.
We disagree with the results in Ref.~\cite{Yu:2017pot} where the UIC contributions are much larger than the coh. ones in the whole $p_{T}$ ranges, and the fragmentation formalism is used beyond its validity scope and the double counting exists.
Thus, the relative contribution of UIC channel to the heavy quarkonium photoproduction increases with the growing quarkonium mass.
And UIC process begins to dominate the photoproduction processes at large $p_{T}$ domain.
Finally, we find that the contributions of photoproduction and fragmentation processes are still non-negligible in bottomonium production.

\section{Summary and conclusion}
\label{Summary and conclusions}

In the framework of the NRQCD, we studied the production of heavy quarkonium originating from inelastic photoproduction and fragmentation processes in $p$-$p$ and $Pb$-$Pb$ collisions  at LHC energies.
We derived the exact treatment by performing a consistent analysis of the terms neglected in transiting from the exact formula to the WWA one,
where the Martin-Ryskin method is adopted to weight the different photon sources, and the BCCKL method is adopted to match the fixed-order and fragmentation contributions.
The full partonic kinematics matched with the exact treatment were also given.
By comprehensively discussing the $Q^{2}$-, $y$-, $z$- and $\sqrt{s}$-dependence behaviours of cross sections,
we obtained the features of WWA in inelastic heavy quarkonium photoproductions in heavy-ion collisions.
In order to estimate the errors existing in the widely employed photon spectra, we calculated the total cross sections.
Meanwhile, the double counting problems and the relative contributions of the ultra-incoherent channel were also discussed.
Finally, we also calculated the $p_{T}$-dependent cross sections to discuss the contributions of photoproduction and fragmentation processes to the inelastic heavy quarkonium productions.

We found that the inelastic photoproduction and fragmentation processes provide valuable contributions to the heavy quarkonium production, especially in the large $p_{T}$ regions.
While the ultra-incoherent photon emission plays very important role in $p$-$p$ collisions, and can also provide the meaningful contribution in $Pb$-$Pb$ collisions;
its relative contribution to inelastic heavy quarkonium photoproductions rapidly increases along with the growing quarkonium mass, and begins to dominate the photoproduction processes at large $p_{T}$ ranges.

The validity scopes of the WWA in the processes discussed in this work are highly restricted, which are compatible with the characteristics of coh. and OIC processes, and possess high accuracy only within the ranges of $Q^{2}<1~\mathrm{GeV}^{2}$, $y<0.1$, $Z\gg1$, and $\sqrt{s}>400$ and $1400~\mathrm{GeV}$ in $p$-$p$ and $Pb$-$Pb$ collisions respectively.
In particular, the WWA has a much higher accuracy in $Pb$-$Pb$ collisions, where the coh. process is enhanced by $Z_{Pb}^{2}$ and the UIC one is greatly suppressed.
Conversely, the kinematical behaviours of the UIC process are contradict to those of WWA, which concentrate on the regions where the WWA errors are prominent.
Therefore, WWA is not a suitable approximation in $p$-$p$ collisions, where the UIC contribution becomes dominant.
Furthermore, we found that the aforementioned equivalent photon spectra are generally derived beyond the applicable scopes of WWA, and the serious double counting exists when different channels are considered simultaneously.
Indeed, the exact treatment is necessary to accurately address the inelastic heavy quarkonium photoproduction in $p$-$p$ and $Pb$-$Pb$ collisions at LHC energies.

It should be noted that, when considering the contribution of photonproduction to the hadronic process, the equivalent photon flux should be integrated to $b_{\textrm{max}}=2R_{A}$.
Integrating to infinity artificially enhances the cross section by including unphysical contributions from non-overlapping nuclei at large.
In present paper, we adopt the empirical approximation to include this truncation affects.
The proper method to incorporate $b$ is the semiclassical impact parameter description (Glauber model), which is more appropriate to calculate the case of heavy ions collisions (recent work, see Ref. ~\cite{Wu:2024jqn}).
In our future work, we will perform our work based on the $b$-dependent scenario.

\section*{ACKNOWLEDGMENTS}

This work is supported in part by the National Natural Science Foundation of China (Grants No.12305092, No.12233006 and No.12150013), by the Xingdian Talent Support Program for Youth Project, and by the Yunnan Fundamental Research Projects (grant NO. 202401AU070206).
H. L. is supported by the Scientific Research and Innovation Project of Postgraduate Students in the Academic Degree of YunNan University (grant NO. KC-242410212).


\appendix
\section{Long-distance matrix elements of heavy quarkonium}
\label{WF_LDMEs}

\begin{table*}[htbp]
\renewcommand\arraystretch{1.7}
\centering
\caption{\label{Kine.inel.Q2.y} The kinematical boundaries of $Q^{2}$, $y$ distributions. $\hat{s}_{\mathrm{min}}=\hat{s}^{*}_{\mathrm{min}}=(m_{T}+p_{T\mathrm{min}})^{2}$, $p_{T}^{2}=\hat{t}(\hat{s}\hat{u}+Q^{2}M_{H}^{2})/(\hat{s}+Q^{2})^{2}$.
}
\begin{tabular}{L{1cm}C{5cm}C{3.5cm}C{4.6cm}C{3cm}}
\hline
\hline
Variables & Coherent direct & UIC direct & Coherent resolved & UIC resolved \\
    \hline
    $z_{\mathrm{min}}$  &  \multicolumn{4}{c}{$\left[(M_{H}^{2}+\hat{s})-\sqrt{(\hat{s}-M_{H}^{2})^{2}-4p_{T \mathrm{min}}^{2}\hat{s}}\right]/(2\hat{s}$)} \\
    $z_{\mathrm{max}}$  &  \multicolumn{4}{c}{$\left[(M_{H}^{2}+\hat{s})+\sqrt{(\hat{s}-M_{H}^{2})^{2}-4p_{T \mathrm{min}}^{2}\hat{s}}\right]/(2\hat{s}$)} \\
    $\hat{t}_{\mathrm{min}}$  &  \multicolumn{2}{c}{$-(1-z_{\mathrm{min}})(\hat{s}+Q^{2})$}  & \multicolumn{2}{c}{$-(1-z_{\mathrm{min}})\hat{s}^{*}$} \\
    $\hat{t}_{\mathrm{max}}$  &  \multicolumn{2}{c}{$-(1-z_{\mathrm{max}})(\hat{s}+Q^{2})$}  & \multicolumn{2}{c}{$-(1-z_{\mathrm{max}})\hat{s}^{*}$} \\
    $z_{a' \mathrm{min}}$    &   $\backslash$   &   $\backslash$   &   $\hat{s}^{*}_{\mathrm{min}}/[y(s_{\alpha b}-m_{\alpha}^{2})]$  &  $\hat{s}^{*}_{\mathrm{min}}/(yx_{a}x_{b}s_{NN})$ \\
    $z_{a' \mathrm{max}}$    &   $\backslash$   &   $\backslash$   &   1   &   1 \\
    $x_{b \mathrm{min}}$    & $(\hat{s}_{\mathrm{min}}+Q^{2})/[y(s_{\alpha b}|_{x_{b\mathrm{max}}}-m_{\alpha}^{2})]$  &  $(\hat{s}_{\mathrm{min}}+Q^{2})/(yx_{a}s_{NN})$  &  $\hat{s}^{*}_{\mathrm{min}}/[y(s_{\alpha b}|_{x_{b\mathrm{max}}}-m_{\alpha}^{2})]$  &  $\hat{s}^{*}_{\mathrm{min}}/(yx_{a}s_{NN})$ \\
    $x_{b \mathrm{max}}$    &  \multicolumn{4}{c}{1}  \\
    $x_{a \mathrm{min}}$    &   $\backslash$   &   $(\hat{s}_{\mathrm{min}}+Q^{2})/(ys_{NN})$   &   $\backslash$   &   $\hat{s}^{*}_{\mathrm{min}}/(ys_{NN})$ \\
    $x_{a \mathrm{max}}$    &   $\backslash$   &   1   &   $\backslash$   &   1 \\
    $y_{\mathrm{min}}$      & \multicolumn{2}{c}{$(\hat{s}_{\mathrm{min}}+Q^{2})/(s_{\alpha b}|_{x_{b\mathrm{max}}}-m_{\alpha}^{2})$} & \multicolumn{2}{c}{$\hat{s}^{*}_{\mathrm{min}}/(s_{\alpha b}|_{x_{b\mathrm{max}}}-m_{\alpha}^{2})$} \\
    $y_{\mathrm{max}}$      & \multicolumn{4}{c}{$\left[\sqrt{Q^{2}(4m_{\alpha}^{2}+Q^{2})}-Q^{2}\right]/(2m_{\alpha}^{2})$} \\
    $Q^{2}_{\mathrm{min}}$   &  \multicolumn{4}{c}{$-2m_{\alpha}^{2}+\left[(s_{\alpha b}+m_{\alpha}^{2})(s_{\alpha b}-\hat{s}+m_{\alpha}^{2})-(s_{\alpha b}-m_{\alpha}^{2})\sqrt{(s_{\alpha b}-\hat{s}+m_{\alpha}^{2})^{2}-4s_{\alpha b}m_{\alpha}^{2}}\right]/(2s_{\alpha b})$} \\
    $Q^{2}_{\mathrm{max}}$   &  \multicolumn{4}{c}{$y(s_{\alpha b}|_{x_{b\mathrm{max}}}-m_{\alpha}^{2})-\hat{s}_{\mathrm{min}}$} \\
\hline
\hline
\end{tabular}
\end{table*}

\begin{table*}[htbp]
\renewcommand\arraystretch{1.7}
\centering
\caption{\label{Kine.inel.pT.z} The kinematical boundaries of $p_{T}$, $z$ distributions.
Where $x_{1}=\hat{s}/s_{\alpha b}$, other boundaries are the same as Table~\ref{Kine.inel.Q2.y}.}
\begin{tabular}{L{1cm}C{4cm}C{4cm}C{4cm}C{4cm}}
\hline
\hline
Variables & Coherent direct & UIC direct & Coherent resolved & UIC resolved \\
    \hline
    $Q^{2}_{\mathrm{min}}$   &  \multicolumn{4}{c}{$x^{2}_{1}m^{2}_{\alpha}/(1-x_{1})$} \\
    $Q^{2}_{\mathrm{max}}$   &  $1/R_{\alpha}^{2}$  & $(1-x_{1})s_{NN}$  & $1/R_{\alpha}^{2}$ & $(1-x_{1})s_{NN}$ \\
    $|y_{r \mathrm{max}}|$  & \multicolumn{4}{c}{$\ln[(\hat{s}_{\mathrm{max}}+M_{H}^{2}+\sqrt{(\hat{s}_{\mathrm{max}}-M_{H}^{2})^{2}
-4p^{2}_{T}\hat{s}_{\mathrm{max}}})/(\hat{s}_{\mathrm{max}}+M_{H}^{2}-\sqrt{(\hat{s}_{\mathrm{max}}-M_{H}^{2})^{2}
-4p^{2}_{T}\hat{s}_{\mathrm{max}}})]^{1/2}$} \\
    $p^{2}_{T\mathrm{min}}$ & \multicolumn{4}{c}{$M^{2}_{H}$} \\
    $p^{2}_{T\mathrm{max}}$ &\multicolumn{4}{c}{$(1-z)\left[z(s_{\alpha b}|_{x_{b\mathrm{max}}}-m_{\alpha}^{2})-M_{H}^{2}\right]$} \\
\hline
\hline
\end{tabular}
\end{table*}
For the reader's convenience and for completeness, we list here the involved LDMEs.
The mass of heavy quarkonia are $M_{J/\psi}=3.097~\mathrm{GeV}$, $M_{\psi(2S)}=3.686~\mathrm{GeV}$, $M_{\Upsilon(1S)}=9.460~\mathrm{GeV}$, $M_{\Upsilon(2S)}=10.023~\mathrm{GeV}$, and $M_{\Upsilon(3S)}=10.355~\mathrm{GeV}$~\cite{Agashe:2014kda}.
The LDMEs of the charmonium are given by~\cite{Butenschoen:2010rq,Sharma:2012dy,Feng:2015wka},
\begin{align}\label{CharmMEs}
&\!\!\!\langle\mathcal{O}^{J/\psi}[^{3}S_{1}^{(1)}]\rangle=1.32~\mathrm{GeV}^{3},\nonumber\\[1mm]
&\!\!\!\langle\mathcal{O}^{J/\psi}[^{1}S_{0}^{(8)}]\rangle=(4.50\pm0.72)\times10^{-2}~\mathrm{GeV}^{3},\displaybreak[0]\nonumber\\
&\!\!\!\langle\mathcal{O}^{J/\psi}[^{3}S_{1}^{(8)}]\rangle=(3.12\pm0.93)\times10^{-3}~\mathrm{GeV}^{3},\displaybreak[0]\nonumber\\%(0.0013\pm0.0013)
&\!\!\!\langle\mathcal{O}^{J/\psi}[^{3}P_{0}^{(8)}]\rangle=(-1.21\pm0.35)\times10^{-2}~\mathrm{GeV}^{5},\displaybreak[0]\\
\displaybreak[0]\nonumber\\
&\!\!\!\langle\mathcal{O}^{\psi(2S)}[^{3}S_{1}^{(1)}]\rangle=0.76~\mathrm{GeV}^{3},\displaybreak[0]\nonumber\\
&\!\!\!\langle\mathcal{O}^{\psi(2S)}[^{1}S_{0}^{(8)}]\rangle=(0.0080\pm0.0067)~\mathrm{GeV}^{3},\displaybreak[0]\nonumber\\
&\!\!\!\langle\mathcal{O}^{\psi(2S)}[^{3}S_{1}^{(8)}]\rangle=(0.00330\pm0.00021)~\mathrm{GeV}^{3},\displaybreak[0]\nonumber\\
&\!\!\!\langle\mathcal{O}^{\psi(2S)}[^{3}P_{0}^{(8)}]\rangle=(0.0080\pm0.0067)m_{c}^{2}~\mathrm{GeV}^{3}.\displaybreak[0]
\end{align}

The LDMEs for the bottomonium~\cite{Feng:2015wka} are,
\begin{align}\label{BottomMEs}
&\!\!\!\langle\mathcal{O}^{\Upsilon(1S)}[^{3}S_{1}^{(1)}]\rangle=9.28~\mathrm{GeV}^{3},\displaybreak[0]\nonumber\\
&\!\!\!\langle\mathcal{O}^{\Upsilon(1S)}[^{1}S_{0}^{(8)}]\rangle=(13.60\pm2.43)\times10^{-2}~\mathrm{GeV}^{3},\displaybreak[0]\nonumber\\%(0.0121\pm0.0400)
&\!\!\!\langle\mathcal{O}^{\Upsilon(1S)}[^{3}S_{1}^{(8)}]\rangle=(0.61\pm0.24)\times10^{-2}~\mathrm{GeV}^{3},\displaybreak[0]\nonumber\\%(0.0477\pm0.0334)
&\!\!\!\langle\mathcal{O}^{\Upsilon(1S)}[^{3}P_{0}^{(8)}]\rangle=(-0.93\pm0.5)m_{b}^{2}\times10^{-2}~\mathrm{GeV}^{3},\displaybreak[0]\\%5m_{b}^{2}\langle\mathcal{O}^{\Upsilon(1S)}[^{1}S_{0}^{(8)}]\rangle
\displaybreak[0]\nonumber\\
&\!\!\!\langle\mathcal{O}^{\Upsilon(2S)}[^{3}S_{1}^{(1)}]\rangle=4.63~\mathrm{GeV}^{3},\displaybreak[0]\nonumber\\
&\!\!\!\langle\mathcal{O}^{\Upsilon(2S)}[^{1}S_{0}^{(8)}]\rangle=(0.62\pm1.98)\times10^{-2}~\mathrm{GeV}^{3},\displaybreak[0]\nonumber\\
&\!\!\!\langle\mathcal{O}^{\Upsilon(2S)}[^{3}S_{1}^{(8)}]\rangle=(2.22\pm0.24)\times10^{-2}~\mathrm{GeV}^{3},\displaybreak[0]\nonumber\\
&\!\!\!\langle\mathcal{O}^{\Upsilon(2S)}[^{3}P_{0}^{(8)}]\rangle=(-0.13\pm0.43)m_{b}^{2}\times10^{-2}~\mathrm{GeV}^{3},\displaybreak[0]\\
\displaybreak[0]\nonumber\\
&\!\!\!\langle\mathcal{O}^{\Upsilon(3S)}[^{3}S_{1}^{(1)}]\rangle=3.54~\mathrm{GeV}^{3},\displaybreak[0]\nonumber\\
&\!\!\!\langle\mathcal{O}^{\Upsilon(3S)}[^{1}S_{0}^{(8)}]\rangle=(1.45\pm1.16)\times10^{-2}~\mathrm{GeV}^{3},\displaybreak[0]\nonumber\\
&\!\!\!\langle\mathcal{O}^{\Upsilon(3S)}[^{3}S_{1}^{(8)}]\rangle=(1.32\pm0.20)\times10^{-2}~\mathrm{GeV}^{3},\displaybreak[0]\nonumber\\
&\!\!\!\langle\mathcal{O}^{\Upsilon(3S)}[^{3}P_{0}^{(8)}]\rangle=(-0.27\pm0.25)m_{b}^{2}\times10^{-2}~\mathrm{GeV}^{3}.\displaybreak[0]
\end{align}
And the multiplicity relations
\begin{align}\label{Multis}
&\langle\mathcal{O}^{H}[^{3}P_{J}^{(8)}]\rangle=(2J+1)\langle\mathcal{O}^{H}[^{3}P_{0}^{(8)}]\rangle,\displaybreak[0]
\end{align}
are used, where $m_{c}$ and $m_{b}$ are the charm quark and bottom quark mass, respectively.

\section{Full kinematical relations}
\label{FKR}

We give here a detailed treatment of the partonic kinematics which is matched with the exact treatment.

The energy and momentum in $\alpha$-$b$ parton level read
\begin{align}\label{Mant.coh.dir}
E_{\alpha}&=\frac{\left(s_{\alpha b}+m_{\alpha}^{2}\right)}{2\sqrt{s_{\alpha b}}},~~E_{b}=\frac{\left(s_{\alpha b}-m_{\alpha}^{2}\right)}{2\sqrt{s_{\alpha b}}},\displaybreak[0]\nonumber\\
p_{\mathrm{CM}}&=\frac{\left(s_{\alpha b}-m_{\alpha}^{2}\right)}{2\sqrt{s_{\alpha b}}},\displaybreak[0]
\end{align}
where the $s_{\alpha b}$ for each photon emission are
\begin{align}\label{s0}
  &s_{\alpha b}|_{\mathrm{coh}.}=m_{A}^{2}+\frac{x_{b}}{N_{B}}\left(s-m_{A}^{2}-m_{B}^{2}\right),\displaybreak[0]\nonumber\\
  &s_{\alpha b}|_{\mathrm{OIC}}=m_{p}^{2}+\frac{x_{b}}{N_{A}N_{B}}\left(s-m_{A}^{2}-m_{B}^{2}\right),\displaybreak[0]\nonumber\\
  &s_{\alpha b}|_{\mathrm{UIC}}=m_{a}^{2}+\frac{x_{a}x_{b}}{N_{A}N_{B}}\left(s-m_{A}^{2}-m_{B}^{2}\right),\displaybreak[0]
\end{align}
where $s=\left(p_{A}+p_{B}\right)^2=\left(N_{A}+N_{B}\right)^{2}s_{NN}/4$ is the energy square in $A$-$B$ process.
While the energy and momentum in $\gamma^{*}$-$b$ parton level are
\begin{align}\label{Epab.}
\hat{E}_{\gamma}&=\frac{\left(\hat{s}-Q^{2}\right)}{2\sqrt{\hat{s}}},~~~~\hat{E}_{H}=\frac{\left(\hat{s}+M_{H}^{2}\right)}{2\sqrt{\hat{s}}},\displaybreak[0]\nonumber\\
\hat{p}_{\mathrm{CM}}&=\frac{\left(\hat{s}+Q^{2}\right)}{2\sqrt{\hat{s}}},~~~\hat{p}_{\mathrm{CM}}'=\frac{\left(\hat{s}-M_{H}^{2}\right)}{2\sqrt{\hat{s}}}.\displaybreak[0]
\end{align}

The Mandelstam variables involved in the case of direct photoproduction are
\begin{align}\label{Mant.}
\hat{s}&=\left(q+p_{b}\right)^{2}=y\left(s_{\alpha b}-m_{\alpha}^{2}\right)-Q^{2},\displaybreak[0]\nonumber\\
\hat{t}&=\left(q-p_{H}\right)^{2}=-\left(1-z\right)\left(\hat{s}+Q^{2}\right),\displaybreak[0]\nonumber\\
\hat{u}&=\left(p_{b}-p_{H}\right)^{2}=M_{H}^{2}-z\left(\hat{s}+Q^{2}\right),\displaybreak[0]
\end{align}
while those for resolved photoproduction are
\begin{align}\label{Mant.}
\hat{s}^{*}&=\left(p_{a'}+p_{b}\right)^{2}=yz_{a'}\left(s_{\alpha b}-m_{\alpha}^{2}\right),\displaybreak[0]\nonumber\\
\hat{t}^{*}&=\left(p_{a'}-p_{H}\right)^{2}=-\left(1-z\right)\hat{s}^{*},\displaybreak[0]\nonumber\\
\hat{u}^{*}&=\left(p_{b}-p_{H}\right)^{2}=M_{H}^{2}-z\hat{s}^{*}.\displaybreak[0]
\end{align}

The kinematical boundaries of the mentioned distributions are given in Tables~\ref{Kine.inel.Q2.y},~\ref{Kine.inel.pT.z}.
When we calculate the contribution of photoproduction to hadronic process in Figs.~\ref{fig:dPT.Charmonium} and \ref{fig:dPT.Bottonium}, the cut $b_{\textrm{max}}=2R_{A}$ should be considered for avoiding the unphysical contributions from non-overlapping nuclei at large.
We adopt the central ideal in Refs.~\cite{Baur:2001jj, CED}, the squared transverse momentum of the photon can be written as

\begin{align}\label{Q_T2}
Q^{2}_{\textrm{T}}&=(1-x)\left(Q^{2}-\frac{x^{2}}{1-x}m_{p}^{2}\right)\sim\frac{1}{b^{2}},\displaybreak[0]\nonumber\\
Q^{2}_{\textrm{min}}&=\frac{b^{-2}_{\textrm{max}}+x^{2}m_{p}^{2}}{1-x}.\displaybreak[0]
\end{align}
where $b_{\textrm{max}}=2R_{A}$ ($R_{A}=A^{1/3}1.2~ \textrm{fm})$.

Finally, we give here the Jacobian determinant $\mathcal{J}$ for each distribution.
In the case of the $Q^{2}$ and $y$ distributions, we have
\begin{align}\label{Jac.Q2.y}
\mathcal{J}=\frac{2r^{2}\left|\textbf{p}_{b}\right|}{E_{\alpha'}E_{b}}=\frac{2r^{2}}{\sqrt{\left(r^{2}+m_{\alpha}^{2}\right)}},\displaybreak[0]
\end{align}
while those for $z$ distribution are
\begin{align}\label{Jac.pT.yr}
\mathcal{J}_{\mathrm{dir.}}=&\frac{x_{b}}{z\left(1-z\right)},~~\mathcal{J}_{\mathrm{res.}}=\frac{z_{a}}{z\left(1-z\right)},\displaybreak[0]
\end{align}
and that for $p_{T}$ distribution is
\begin{align}\label{Jac.pT.yr}
\mathcal{J}=&\frac{\left(\hat{s}^{3/2}+Q^{2}\sqrt{\hat{s}}\right)}{y\left(s_{\alpha b}|_{x_{b\mathrm{max}}}-m_{\alpha}^{2}\right)\left(\sqrt{\hat{s}}-\cosh y_{r}m_{T}\right)},\displaybreak[0]
\end{align}
for coherent-direct process.
And the relations between Eq.~({\ref{Jac.pT.yr}}) and the rest cases are: $\mathcal{J}_{\mathrm{incoh.dir.}}=\mathcal{J}/x_{a}$, $\mathcal{J}_{\mathrm{coh.res.}}=\mathcal{J}/x_{b}$, and
$\mathcal{J}_{\mathrm{incoh.res.}}=\mathcal{J}/x_{a}x_{b}$.
In the case of fragmentation processes, $\mathcal{J}=\left(\hat{s}+Q^{2}\right)/\left(\cosh y_{r}\sqrt{\hat{s}}\right)$.

\end{document}